\documentclass[11pt,a4paper]{article}
\pdfoutput=1
\usepackage{jheppub}
\usepackage{bm}

\usepackage{url}

\allowdisplaybreaks[2]

\def\cN{\mathcal{N}}

\def\cO{\mathcal{O}}
\def\cJ{\mathcal{J}}
\def\cD{\mathcal{D}}

\def\cC{\mathcal{C}}
\def\cS{\mathcal{S}}
\def\cI{\mathcal{I}}
\def\mint{\int_{-\infty}^\infty\!\cdots\!\int_{-\infty}^\infty}

\newcommand{\be}{\begin{equation}}
\newcommand{\ee}{\end{equation}}
\newcommand{\ba}{\begin{aligned}}
\newcommand{\ea}{\end{aligned}}

\def\Res{\mathop {\rm Res} \limits}

\def\bra#1{\left\langle #1 \right|}
\def\ket#1{\left| #1 \right\rangle}

\DeclareMathOperator{\Li}{Li}
\DeclareMathOperator{\Ai}{Ai}

\def\({\left(}
\def\){\right)}

\newcommand{\pd}{\partial}
\DeclareMathOperator{\real}{Re}

\DeclareMathOperator{\Tr}{Tr}
\DeclareMathOperator{\Det}{Det}

\DeclareMathOperator{\B}{B}

\newcommand{\re}{{\rm e}}
\newcommand{\ri}{{\rm i}}
\newcommand{\rd}{{\rm d}}

\newcommand{\mx}{\mathsf{x}}

\newcommand{\mm}{\mathsf{p}}

\title{ABJM on ellipsoid and topological strings}

\author{Yasuyuki Hatsuda}

\affiliation{D\'epartement de Physique Th\'eorique et Section de Math\'ematiques\\
Universit\'e de Gen\`eve, Gen\`eve, CH-1211 Switzerland}

\emailAdd{Yasuyuki.Hatsuda@unige.ch}

\abstract{
It is known that the large $N$ expansion of the partition function in ABJM theory on a three-sphere is completely determined 
by the topological string on local Hirzebruch surface $\mathbb{F}_0$.
In this note, we investigate the ABJM partition function on an ellipsoid, which has a conventional deformation parameter $b$.
Using 3d mirror symmetry, we find a
remarkable relation between the ellipsoid partition function for $b^2=3$ (or $b^2=1/3$) in ABJM theory at $k=1$ and
a matrix model for the topological string on another Calabi-Yau threefold, known as local $\mathbb{P}^2$.
As in the case of $b=1$, we can compute the full large $N$ expansion of the partition function in this case.
This is the first example of the complete large $N$ solution in ABJM theory on the squashed sphere.
Using the obtained results, we also analyze the supersymmetric R\'enyi entropy.
}

\begin{document}

\maketitle

\renewcommand{\thefootnote}{\arabic{footnote}}
\setcounter{footnote}{0}
\setcounter{section}{0}

\section{Introduction}
Supersymmetric gauge theories provide us many important insights and exact results.
Supersymmetric localization is now a basic tool to compute a class of observables exactly.
In this way, path integrals reduce to matrix integrals \cite{Pestun}, and one can explore their large $N$ expansions
quantitatively.
From holographic point of view, these provide predictions of dual gravity theories.

Here we focus on the 3d $\cN=6$ superconformal Chern-Simons-matter theory with quiver
gauge group $U(N)_k \times U(N)_{-k}$, well-known as ABJM theory \cite{ABJM, ABJ}. 
This theory is a fundamental theory on multiple M2-branes.
As shown in \cite{KWY1} (see also \cite{Jafferis, HHL1}), 
the partition function in ABJM theory (and in wider Chern-Simons-matter theories) 
on a three-sphere reduces to a matrix integral by localization.
An interesting observation in \cite{MP-top} is that the ABJM matrix model is closely related to the Chern-Simons matrix model on a lens space $L(2,1)$.
It is known that this lens space matrix model is large $N$ dual to the topological string on local Hirzebruch surface $\mathbb{F}_0$ \cite{AKMV}.
As a result of this chain, the large $N$ expansion in the ABJM matrix model is expected to be captured by the topological string on local $\mathbb{F}_0$.
In fact, the all-genus free energy in the 't Hooft limit ($N \to \infty$ with fixed $N/k$) was computed in \cite{DMP1} 
by the holomorphic anomaly equation in the topological strings.
It is also important to note that the direct saddle-point analysis in the M-theory limit ($N \to \infty$ with fixed $k$) revealed the expected $N^{3/2}$ 
behavior \cite{HKPT}.
This result is a piece of evidence that ABJM theory is indeed a theory on multiple M2-branes.
Surprisingly, the complete large $N$ expansion, including all non-perturbative corrections in $1/N$,
turned out to be determined by the topological string on local $\mathbb{F}_0$ in a highly non-trivial way \cite{HMMO}.
The result heavily relies on a formulation in \cite{MP}, called the Fermi-gas formalism (see \cite{HMO-review} for
developments on the Fermi-gas formalism).
In summary, the large $N$ problem in ABJM theory on the three-sphere has been solved with the help of topological string theory.

In this note, we initiate an investigation of the large $N$ expansion in ABJM theory on a squashed three-sphere preserving $U(1) \times U(1)$
isometry, known as an ellipsoid.
The 3d ellipsoid has a conventional deformation parameter $b$, defined in \eqref{eq:b-def}.
In the limit $b\to 1$, it reduces to a round sphere.
There are several motivations to consider theories on the ellipsoid.
Firstly, ellipsoid partition functions have a factorization property into vortex and anti-vortex partition functions, 
as shown in \cite{Pasquetti, BDP}.
To see this property, one needs to turn on the parameter $b \ne 1$, because for $b=1$, the vortex and anti-vortex
contributions are not separable.
Secondly, 3d ellipsoid partition functions are related to open topological strings \cite{Pasquetti, LV}.%
\footnote{The appearance of topological strings in this note is completely different from this context.
We find a novel relation between the (mirror) ABJM matrix model and a matrix model for the \textit{closed} topological string
on particular Calabi-Yau.
This relation is probably accidental.} 
The squashing parameter is identified as the string coupling of the open topological strings.
The ellipsoid partition function is expected to provide a kind of non-perturbative completion of the topological strings.   
Thirdly, the ellipsoid partition functions are related to a supersymmetric generalization of the R\'enyi entropy \cite{NY}.
The curious fact in \cite{NY} is that the partition functions in $\cN=2$ supersymmetric gauge theories on branched spheres 
are equivalent to those on ellipsoids.
As a result, one can compute the supersymmetric R\'enyi entropy by the ellipsoid partition functions.
From these examples, it is desirable to understand the large $N$ behavior of the ellipsoid partition functions more deeply.

The partition functions in general $\cN=2$ supersymmetric gauge theories on squashed three-spheres 
were computed in \cite{HHL2, IY} by using localization.
One can easily apply their results to ABJM theory on the ellipsoid with squashing parameter $b$, and obtains the following matrix integral:
\be
\ba
Z_{b^2}^{\rm ABJM}(k,N)&=\frac{1}{(N!)^2} \int \rd^N \sigma \rd^N \tilde{\sigma} \,
\re^{\pi \ri k \sum_{i} (\sigma_i^2-\tilde{\sigma}_i^2)} 
\prod_{i,j} \frac{s_b(\sigma_i-\tilde{\sigma}_j+\frac{\ri Q}{4})^2}{s_b(\sigma_i-\tilde{\sigma}_j-\frac{\ri Q}{4})^2} \\
&\qquad \times \prod_{i<j} 4 \sinh(\pi b \sigma_{ij} )\sinh(\pi b^{-1} \sigma_{ij} )\cdot 
4 \sinh(\pi b \tilde{\sigma}_{ij} )\sinh(\pi b^{-1} \tilde{\sigma}_{ij} ),
\ea
\label{eq:Z-ABJM}
\ee
where $Q=b+1/b$, and $s_b(z)$ is the double sine function defined by \eqref{eq:double-sine}.
We are using compact notations $\sigma_{ij}=\sigma_i-\sigma_j$ and $\tilde{\sigma}_{ij}=\tilde{\sigma}_i-\tilde{\sigma}_j$.
In the round sphere limit $b \to 1$, the ratio of the double sine function in the integrand reduces to the hyperbolic function,
and the resulting matrix model reproduces the original result in \cite{KWY1}.
Obviously, the matrix model \eqref{eq:Z-ABJM} is much more complicated than 
the original matrix model for $b=1$.
The analysis at large $N$ is quite limited so far \cite{IY, MPS}.

One approach to analyze the ABJM matrix model is to use 3d mirror symmetry \cite{IS, HaWi}.
It is known that ABJM theory at $k=1$ is dual to the $\mathcal{N}=4$
$U(N)$ SYM with an adjoint hypermultiplet and a fundamental hypermultiplet.
In this note, we refer to the latter as the mirror theory, for short.
The partition function of the mirror theory is also computed by localization
\be
\ba
Z_{b^2}^\text{Mirror}(N)&=\frac{1}{N!} \int \! \rd^N \lambda 
\prod_{i} \frac{s_b(\lambda_{i}+\frac{\ri Q}{4})}{s_b(\lambda_{i}-\frac{\ri Q}{4})}
\prod_{i,j} \frac{s_b(\lambda_{ij}+\frac{\ri Q}{4})}{s_b(\lambda_{ij}-\frac{\ri Q}{4})} \\
&\qquad \times \prod_{i<j} 4\sinh (\pi b \lambda_{ij})\sinh (\pi b^{-1} \lambda_{ij} ).
\ea
\label{eq:Z-mirror}
\ee
Mirror symmetry predicts that these two partition functions are \textit{exactly} equivalent,
\be
Z_{b^2}^\text{ABJM}(k=1,N)=Z_{b^2}^\text{Mirror}(N),
\label{eq:mirror-equal-k=1}
\ee
for \textit{any} $b$ and $N$.
This equality for $b=1$ was proved in \cite{KWY2} (see also \cite{DF} for another interesting perspective). 
The crucial idea in their proof is to use the Cauchy determinant formula.
Since, for general $b$, it seems that one cannot use this formula easily, a proof of \eqref{eq:mirror-equal-k=1} is challenging.
As shown in the next section, we can confirm it for $N=1,2$ by evaluating the matrix integrals. 
In this note, we assume \eqref{eq:mirror-equal-k=1} for general $b$ and $N$.
Though the mirror partition function \eqref{eq:Z-mirror} looks simpler than the ABJM partition function \eqref{eq:Z-ABJM} at $k=1$,
it is still difficult to extract information at large $N$.

As a first step to analyze the matrix model \eqref{eq:Z-mirror}, we start by looking for special cases, in which the matrix model
simplifies, as in $b=1$.
We find that such a simplification indeed happens if $b^2$ is odd.%
\footnote{We thank Masazumi Honda for telling us that the simplification occurs not only for $b^3=1,3$ but
also for odd $b^2$.}
In these cases, the ratio of the double sine function again reduces to the hyperbolic functions.
In particular, in the special case $b^2=3$ (or equivalently $b^2=1/3$), the matrix model drastically simplifies.
Quite remarkably, we find that in this case there is a non-trivial relation to the topological string on local $\mathbb{P}^2$.
Before closing this section, let us briefly state this fact.
In \cite{MZ, KMZ}, a new class of matrix models for topological strings was constructed.
Their proposal is based on a quantum mechanical reformulation of the topological strings in \cite{GHM1, KM}.
In particular, the matrix model corresponding to local $\mathbb{P}^2$ is given by
\be
Z_{\mathbb{P}^2}(\hbar,N)=\frac{1}{N!} \int \! \frac{\rd^N p}{b^N} \prod_i | \Psi_{a,c}(p_i)|^2
\frac{\prod_{i<j}4 \sinh^2(\frac{\pi}{b} (p_i-p_j) )}{\prod_{i,j} 2\cosh (\frac{\pi}{b}(p_i-p_j)+\frac{\pi \ri}{6} )},
\quad \hbar=\frac{2\pi b^2}{3},
\label{eq:Z-P2}
\ee
where $| \Psi_{a,c}(p_i)|^2$ is represented by the ratio of the double sine function, as in \eqref{eq:Psi_ac}.
The main claim in \cite{MZ} is that the 't Hooft expansion ($N \to \infty$ with $N/\hbar$ fixed) of this matrix model
describes the (unrefined) topological string on local $\mathbb{P}^2$ with string coupling $g_s=1/\hbar$.
This is a natural consequence from the result in \cite{GHM1}.%
\footnote{In the semi-classical regime $\hbar \to 0$, the system is governed by the refined topological strings
in the Nekrasov-Shatashvili limit \cite{NS}, called quantum geometry \cite{ACDKV, HKRS}. 
The fact that the same quantum mechanical system \textit{at strong coupling} $\hbar \to \infty$
describes the unrefined topological strings with string coupling $g_s=1/\hbar$ is highly non-trivial and surprising.
This is one of the main conjectures in \cite{GHM1}, and has been confirmed in many examples \cite{MZ, KMZ, GKMR, CGM2}.
This approach is also powerful in solving a wide class of relativistic integrable systems \cite{HM, FHM-cluster}.}
As shown in Section~\ref{sec:b3}, the matrix model \eqref{eq:Z-P2} in the case of $b^2=3$ ($\hbar=2\pi$) exactly coincides with the
mirror matrix model \eqref{eq:Z-mirror} with the same $b$!
We conclude that in the case of $b^2=3$, the following triality relation holds for any $N$:
\be
Z_{b^2=3}^\text{ABJM}(k=1,N)=Z_{b^2=3}^\text{Mirror}(N)=Z_{\mathbb{P}^2}(\hbar=2\pi, N).
\label{eq:triality-b3}
\ee
Note again that the first equality is currently an assumption to be proved, while the second equality is exactly true.
The physical reason of this relation is mysterious. 
In our analysis, we need the large $N$ expansion not in the 't Hooft limit but in the M-theoretic limit: $N \to \infty$ with $\hbar=2\pi$.
This case has already been studied in \cite{GHM1} in great detail, and it turned out that a generating function
of \eqref{eq:Z-P2} can be written in closed form, as in \eqref{eq:Xi-b3}.
As a consequence,
we can know the complete large $N$ expansion of the $b^2=3$ ellipsoid partition function
with the help of topological string theory (see \eqref{eq:Z3-largeN}).
The leading $N^{3/2}$ behavior is in perfect agreement with the known results in \cite{IY, MPS}.
It is fantastic that the ellipsoid partition function in ABJM theory for $b^2=1,3$ are both determined by topological string theory
on \textit{different} Calabi-Yau threefolds.
Once the ellipsoid partition function is known, the supersymmetric R\'enyi entropy
is easily computed.
We present several new results on the supersymmetric R\'enyi entropy in ABJM and its mirror.

The organization of this note is the following.
In Section~2, we start by reviewing the ellipsoid partition function in ABJM theory.
We also consider 3d mirror symmetry, and give non-trivial evidence.
Section~3 is the main part in this note. We show that in the special case $b^2=3$,
the matrix model simplifies, and it coincides with the matrix model for the topological string
on local $\mathbb{P}^2$ proposed by Mari\~no and Zakany.
This remarkable connection allows us to compute the large $N$ expansion including all the non-perturbative
corrections.
We also present a simple generalization of the mirror theory.
In Section~4, using the results in the previous section, we analyze the supersymmetric R\'enyi entropy.
We discuss its large $N$ expansion.
Section~5 is devoted to concluding remarks.
In Appendix~A, some important properties of the double sine function and its related functions are summarized.
In Appendix~B, explicit computations of the ABJM and mirror matrix models for $N=1,2$ are shown.
In Appendix~C, we give a summary on the free energy of the topological string on local $\mathbb{P}^2$,
which is useful to compute the large $N$ expansion of the ellipsoid partition function for $b^2=3$.

\section{ABJM on ellipsoid and 3d mirror symmetry}
\subsection{The ellipsoid partition function}
In this note, we investigate the ellipsoid partition function in ABJM theory.
A three-dimensional ellipsoid can be embedded into $(x_1,x_2,x_3,x_4) \in \mathbb{R}^4$ by
\be
\omega_1^2(x_1^2+x_2^2)+\omega_2^2(x_3^2+x_4^2)=1.
\ee 
Obviously it preserves only the $U(1) \times U(1)$ isometry.
Since the dependence of $\omega_1$ and $\omega_2$ always appears as their ratio,
we introduce parameters
\be
b^2:=\frac{\omega_1}{\omega_2},\qquad Q:=b+b^{-1}.
\label{eq:b-def}
\ee
Partition functions in ${\cal N}=2$ supersymmetric theories on the ellipsoid were
computed in \cite{HHL2} by using localization.
Since ABJM theory is the supersymmetric Chern-Simons-matter theory with gauge group $U(N)_k \times U(N)_{-k}$,
where $k$ is the Chern-Simons level, one can easily write down its ellipsoid partition function.
The theory has four bi-fundamental chiral multiplets.
Two of them belong to the $(N,\overline{N})$ representation, while the other two belong to the $(\overline{N},N)$ representation.
All these chiral multiplets have the Wyle weight $\Delta=1/2$.

The localization technique allows us to compute the partition function exactly.
Using the general formula in \cite{HHL2}, the partition function in ABJM theory reduces to the following matrix model
\be
\ba
Z_{b^2}^{\rm ABJM}(k,N)=\frac{1}{(N!)^2} \int \rd^N \sigma \rd^N \tilde{\sigma} \,
\re^{\pi \ri k \sum_{i} (\sigma_i^2-\tilde{\sigma}_i^2)} 
Z_{b^2}^\text{vec} Z_{b^2}^\text{bi-fund}
\ea
\label{eq:Z-ABJM-ellipsoid}
\ee
where
the contributions of the vector multiplets and the bi-fundamental chiral multiplets are
\be
\ba
Z_{b^2}^\text{vec}&= \prod_{i<j} 4 \sinh(\pi b \sigma_{ij} )\sinh(\pi b^{-1} \sigma_{ij} )\cdot 
4 \sinh(\pi b \tilde{\sigma}_{ij} )\sinh(\pi b^{-1} \tilde{\sigma}_{ij} ) ,\\
Z_{b^2}^\text{bi-fund}&= \prod_{i,j} s_b\(\frac{\ri Q}{4}+\sigma_i-\tilde{\sigma}_j \)^2
s_b \( \frac{\ri Q}{4}-\sigma_i+\tilde{\sigma}_j \)^2
=\prod_{i,j} \frac{s_b(\sigma_i-\tilde{\sigma}_j+\frac{\ri Q}{4})^2}{s_b(\sigma_i-\tilde{\sigma}_j-\frac{\ri Q}{4})^2}.
\ea
\ee
The indices $i,j$ run from $1$ to $N$.
Throughout this note, the double sine function always appears as the ratio,
and it is very useful to define a new function by
\be
\cD_b(\lambda):=\frac{s_b(\lambda+\frac{\ri Q}{4})}{s_b(\lambda-\frac{\ri Q}{4})}.
\label{eq:Db-def}
\ee
Then, the bi-fundamental part is simply written as
\be
Z_{b^2}^\text{bi-fund} = \prod_{i,j} \cD_b(\sigma_i-\tilde{\sigma}_j)^2.
\ee
The function $\cD_b(\lambda)$ has several nice properties.
Some basic properties of $s_b(z)$ and $\cD_b(\lambda)$ are summarized in Appendix~\ref{sec:double-sine}.
Since the function $\cD_b(\lambda)$ is symmetric under $b \leftrightarrow b^{-1}$, 
the partition function also has this symmetry. One can assume $b \geq 1$ without loss of generality.

The round sphere limit $b=1$ is a self-dual point.
In this limit, each contribution becomes
\be
\ba
Z_{b^2=1}^\text{vec}&=\prod_{i<j} \(2 \sinh \pi \sigma_{ij} \)^2 \(2 \sinh \pi \tilde{\sigma}_{ij} \)^2, \\
Z_{b^2=1}^\text{bi-fund}&=
\prod_{i,j} \frac{1}{(2\cosh \pi (\sigma_i -\tilde{\sigma}_j))^2},
\ea
\ee
and the original result \cite{KWY1} is recovered.

Though the partition function exactly reduces to the finite dimensional matrix model \eqref{eq:Z-ABJM-ellipsoid},
the analysis at large $N$ (and also at finite $N$) is still highly non-trivial.
The leading $N^{3/2}$ behavior in the M-theory limit ($N \to \infty$ with fixed $k$) was confirmed in \cite{IY, MPS} based on the analysis in \cite{HKPT}.
In the round sphere case, the complete large $N$ expansion has been known with the help of the topological string 
on local $\mathbb{F}_0$ \cite{DMP1, HMMO}.

\subsection{3d mirror symmetry}
One interesting purpose to study partition functions is to see exact dualities.
It is well-known that ABJM theory at $k=1$ is dual to the $\mathcal{N}=4$
$U(N)$ SYM with an adjoint hypermultiplet and a fundamental hypermultiplet.
This is a kind of 3d mirror symmetry.
The partition function of the mirror theory is also computed by localization
\be
\ba
Z_{b^2}^\text{Mirror}(N)=\frac{1}{N!} \int \! \rd^N \lambda \, \widetilde{Z}_{b^2}^\text{vec}
\widetilde{Z}_{b^2}^\text{adj} \widetilde{Z}_{b^2}^\text{fund},
\ea
\label{eq:Z-mirror-k=1}
\ee
where
\be
\ba
\widetilde{Z}_{b^2}^\text{vec}&=\prod_{i<j} 4\sinh (\pi b \lambda_{ij})\sinh (\pi b^{-1} \lambda_{ij} ),\\
\widetilde{Z}_{b^2}^\text{adj}&=\prod_{i,j} s_b\( \frac{\ri Q}{4}+\lambda_{ij} \)s_b\( \frac{\ri Q}{4}-\lambda_{ij} \)
=\prod_{i,j} \cD_b(\lambda_{ij}) ,\\
\widetilde{Z}_{b^2}^\text{fund}&= \prod_i s_b\( \frac{\ri Q}{4}+\lambda_{i} \)s_b\( \frac{\ri Q}{4}-\lambda_{i} \)
=\prod_{i} \cD_b(\lambda_{i} ).
\ea
\ee
Mirror symmetry states that the two partition functions \eqref{eq:Z-ABJM-ellipsoid} and \eqref{eq:Z-mirror-k=1} 
should be exactly equal, as in \eqref{eq:mirror-equal-k=1}.
Though we do not have a proof of the equality \eqref{eq:mirror-equal-k=1} for arbitrary $b$ and $N$,
we can check it for $N=1,2$.
The detail of the explicit computations is presented in Appendix~\ref{sec:explicit}.
For $N=1$, we can exactly perform the integral, and get the same result on the both sides:
\be
Z_{b^2}^\text{ABJM}(1,1)=Z_{b^2}^\text{Mirror}(1)=\cD_b(0)^2=s_b \( \frac{\ri Q}{4} \)^4 .
\ee
For $N=2$, we also find the following representations:
\be
\ba
Z_{b^2}^\text{ABJM}(k,2)&=\frac{1}{4k}\int \! \rd x \rd y\, \re^{2\pi \ri k x y}\, \cD_b(x)^4 \cD_b(y)^4 \\
&\hspace{-1.7cm}\times 4\sinh(\pi b (x+y))\sinh(\pi b^{-1} (x+y))\cdot 4\sinh(\pi b (x-y)\sinh(\pi b^{-1}(x-y)),
\ea
\label{eq:ZABJM-2}
\ee
and
\be
\ba
Z_{b^2}^\text{Mirror}(2)&=\frac{\cD_b(0)^2}{2} \int \! \rd x \rd y \, \re^{2\pi \ri x y}\,
\cD_b(x)^2 \cD_b(y)^2 \cdot 4\sinh (\pi b y) \sinh (\pi b^{-1} y).
\ea
\label{eq:Zmirror-2}
\ee
These two integrals still look quite different.
To test the equality \eqref{eq:mirror-equal-k=1} for $N=2$, we evaluate these integrals numerically. 
In Table~\ref{tab:mirror-test}, the numerical values for various $b$
are shown. These two partition functions indeed give the same values.
In the remaining sections, we assume 3d mirror symmetry \eqref{eq:mirror-equal-k=1} for general $N$.

\begin{table}[tb]
\caption{A test of 3d mirror symmetry for $N=2$. We evaluate the two integrals \eqref{eq:ZABJM-2} and \eqref{eq:Zmirror-2} independently,
and confirm that they lead to the same values.}
\label{tab:mirror-test}
\begin{center}
  \begin{tabular}{ccc}\hline
$b^2$ & $-\log Z_{b^2}^\text{ABJM}(1,2)$ & $-\log Z_{b^2}^\text{Mirror}(2)$  \\ \hline  
$1$    & $3.9173186080892$  &  $3.9173186080892$  \\
$2$    & $4.6341504360497$  &  $4.6341504360497$  \\
$3$    & $5.8195258638936$  &  $5.8195258638936$  \\
$4$    & $7.1149342370026$  &  $7.1149342370026$  \\
$5$    & $8.4516770209410$  &  $8.4516770209410$  \\
\hline
\end{tabular}
\end{center}
\end{table}

\section{The special case $b=\sqrt{3}$}\label{sec:b3}
\subsection{Relation to the topological string on local $\mathbb{P}^2$}
In general, the ellipsoid partition functions \eqref{eq:Z-ABJM-ellipsoid} and \eqref{eq:Z-mirror-k=1}
are written in terms of the double sine function, and their evaluation is complicated even for very small $N$.
In the round-sphere case $b=1$, the matrix models are expressed in terms of the hyperbolic functions.
In this special case, one can analyze the matrix models both analytically and numerically. 
It is natural to look for some other special values of $b$, for which the matrix models also simplify.
We find that such a simplification indeed happens in the case that $b^2$ is odd.
In particular, in the case of $b^2=3$, the matrix model drastically simplifies.
A key observation is the reduction of $\cD_{\sqrt{3}}(\lambda)$, as in \eqref{eq:Db-sqrt3}.
Using this equation, one immediately obtains
\be
\ba
Z_{b^2=3}^\text{ABJM}(k,N)&=\frac{1}{(N!)^2}\int \! \rd^N \sigma \rd^N \tilde{\sigma}
\, \re^{\pi \ri k\sum_{i} (\sigma_i^2- \tilde{\sigma}_i^2)}\prod_{i,j} \frac{\sinh^2 (\frac{\pi}{\sqrt{3}}(\sigma_i-\tilde{\sigma}_j))}
{\sinh^2 (\sqrt{3}\pi(\sigma_i-\tilde{\sigma}_j))} \\
&\times\prod_{i<j} 4\sinh (\sqrt{3} \pi \sigma_{ij} )\sinh \(\frac{\pi}{\sqrt{3}} \sigma_{ij} \)
\cdot 4\sinh (\sqrt{3} \pi \tilde{\sigma}_{ij} )\sinh \(\frac{\pi}{\sqrt{3}} \tilde{\sigma}_{ij} \) .
\ea
\ee
Similarly the mirror matrix model \eqref{eq:Z-mirror-k=1} reduces to
\be
\ba
Z_{b^2=3}^\text{Mirror}(N) = \frac{1}{N!}
\int\! \frac{\rd^N \lambda}{3^N } \prod_i \frac{\sinh (\frac{\pi}{\sqrt{3}}\lambda_i)}{\sinh (\sqrt{3}\pi \lambda_i)}
 \prod_{i<j} \frac{4 \sinh^3 (\frac{\pi}{\sqrt{3}}\lambda_{ij})}{\sinh (\sqrt{3}\pi \lambda_{ij})},
\ea
\label{eq:Z-mirror-sqrt3-0}
\ee
where we have used an identity:
\be
\prod_{i,j} \cD_b(\lambda_{ij})=\cD_b(0)^N \prod_{i<j} \cD_b(\lambda_{ij})^2.
\ee

Now we see a novel relation between the matrix model \eqref{eq:Z-mirror-sqrt3-0}
and a matrix model proposed in \cite{MZ}.
In \cite{MZ, KMZ}, new matrix models for topological strings were proposed,
based on the earlier results \cite{GHM1, KM}.
In particular, the matrix model corresponding to local $\mathbb{P}^2$ is given by \eqref{eq:Z-P2}.
The function $|\Psi_{a,c}(p)|^2$ is given by
\be
|\Psi_{a,c}(p)|^2=\re^{2\pi (a-c) p} \frac{s_b(p+\ri(a+c))}{s_b(p-\ri (a+c))}, \qquad
a=\frac{b}{2}-\frac{1}{b},\qquad c=\frac{1}{2b}.
\label{eq:Psi_ac}
\ee
In this context, the parameter $b$ is related to the Planck constant $\hbar$, as in \eqref{eq:Z-P2}.
See \cite{MZ, KMZ} for more detail.
Now we set $b=\sqrt{3}$ and thus $\hbar=2\pi$.
Then \eqref{eq:Psi_ac} becomes
\be
|\Psi_{a,c}(p)|^2=\frac{s_{\sqrt{3}}(p+\frac{\ri}{\sqrt{3}})}{s_{\sqrt{3}}(p-\frac{\ri}{\sqrt{3}})}
=\cD_{\sqrt{3}}(p)
=\frac{\sinh ( \frac{\pi}{\sqrt{3}} p)}{\sinh ( \sqrt{3}\pi p)},\qquad
a=c=\frac{1}{2\sqrt{3}}.
\ee
Using an identity
\be
\prod_{i,j} \frac{1}{2\cosh(\frac{\pi}{\sqrt{3}}p+\frac{\pi \ri}{6})}
=\frac{1}{(\sqrt{3})^N} \prod_{i<j} \frac{\sinh (\frac{\pi}{\sqrt{3}}p)}{\sinh (\sqrt{3} \pi p)},
\ee
we conclude that the equality \eqref{eq:triality-b3} exactly holds for any $N$

\subsection{An ideal quantum Fermi-gas and the large $N$ expansion}
Clearly, the matrix model \eqref{eq:Z-mirror-sqrt3-0} can be regarded as a partition function
of an $N$-particle \textit{interacting classical gas} with an external source.
This is in general true for the mirror partition function \eqref{eq:Z-mirror}.
The highly non-trivial statement in \cite{MP} is that this partition function also can be
interpreted as a partition function of an $N$-particle \textit{non-interacting quantum Fermi-gas}
(see also \cite{KKN} for the very similar structure in another matrix model).
This picture does not seem to work in the general case \eqref{eq:Z-mirror}. 
It is allowed only for restricted cases ($b^2=1,3$, for example).%
\footnote{However, the possibility that other values of $b$ admit the ideal Fermi-gas description has not been ruled out.
We have not found these values so far, but they might perhaps exist. 
In general, it is possible to rewrite the partition function as the form of an $N$-particle \textit{interacting quantum Fermi-gas},
as in \cite{MP2}.} 
Following the argument \cite{MP}, we can easily go to the Fermi-gas formalism.%
We first rewrite the partition function \eqref{eq:Z-mirror-sqrt3-0},
by rescaling the integration variables $x_i=2\pi \lambda_i/\sqrt{3}$, as
\be
Z_{b^2=3}(N) =\frac{1}{N!}
\int\! \frac{\rd^N x}{(2\pi \sqrt{3})^N} \prod_i \frac{\sinh (\frac{x_i}{2})}{\sinh (\frac{3x_i}{2})}
 \prod_{i<j} \frac{4 \sinh^3 (\frac{x_i-x_j}{2})}{\sinh (\frac{3(x_i-x_j)}{2})}.
\label{eq:Z-mirror-sqrt3}
\ee
where we have omitted the subscript ``Mirror'' for simplicity.
One can further rewrite it, as in \eqref{eq:Z-P2},
\be
Z_{b^2=3}(N)=\frac{1}{N!}\int\! \frac{\rd^N x}{(2\pi)^N} \prod_{i} \frac{\sinh(\frac{x_i}{2})}{\sinh(\frac{3x_i}{2})}
\frac{\prod_{i<j} 4\sinh^2 (\frac{x_i-x_j}{2})}{\prod_{i,j} 2\cosh ( \frac{x_i-x_j}{2}+\frac{\pi \ri}{6})}.
\ee
Then, after using the Cauchy determinant formula,
the partition function is written as
\be
Z_{b^2=3}(N)=\frac{1}{N!}\sum_{\sigma \in S_N} (-1)^\sigma
\int \! \rd^N x \prod_{i} \rho(x_i, x_{\sigma(i)}),
\label{eq:Z-FG}
\ee
where
\be
\rho(x_1,x_2)=\frac{1}{2\pi} \( \frac{\sinh(\frac{x_1}{2})}{\sinh(\frac{3x_1}{2})} \)^{1/2}
 \frac{1}{2\cosh (\frac{x_1-x_2}{2}+\frac{\pi \ri}{6})} \( \frac{\sinh(\frac{x_2}{2})}{\sinh(\frac{3x_2}{2})} \)^{1/2}.
 \label{eq:rho}
\ee
This density matrix indeed agrees with the one for local $\mathbb{P}^2$ in \cite{KM} (see also \cite{OZ}) for $\hbar=2\pi$.
Note that the density matrix is self-adjoint:
\be 
\rho^\dagger(x_1,x_2)=\overline{\rho(x_2,x_1)}=\rho(x_1,x_2).
\ee
Therefore its eigenvalues are real (and positive).
In the analysis at large $N$, it is convenient to go to the grand canonical ensemble
\be
\Xi_{b^2}(\mu):=1+\sum_{N=1}^\infty \re^{\mu N} Z_{b^2}(N).
\ee
Then the grand canonical partition function for \eqref{eq:Z-FG} is written as that for an ideal quantum Fermi-gas
\be
\Xi_{b^2=3}(\mu)=\Det(1+\re^{\mu} \hat{\rho})=\prod_{n=0}^\infty (1+\re^{\mu-E_n} ),
\ee
where $\Det$ means the Fredholm determinant for the operator $\hat{\rho}$.
In the current case, the Planck constant is set to be $\hbar=2\pi$. 
The one-particle eigenvalue problem in this Fermi-gas system is not the standard Schr\"odinger equation
but rather a Fredholm integral equation for the integral kernel \eqref{eq:rho}
\be
\int_{-\infty}^\infty \rd x'\, \rho(x,x') \phi_n (x')=\re^{-E_n} \phi_n(x),\qquad n=0,1,2,\dots.
\ee
Remarkably this eigenvalue problem was solved in \cite{GHM1} for any $\hbar$,
and the resulting exact quantization condition enjoys a beautiful S-dual structure \cite{WZH} 
(see also \cite{HW, Hatsuda2}).
The grand partition function is also written as
\be
\Xi_{b^2=3}(\mu)=\exp \left[ -\sum_{\ell=1}^\infty \frac{(-\re^{\mu})^\ell}{\ell} \Tr \rho^\ell \right].
\ee
The exact values of $\Tr \rho^\ell$ for the very first few $\ell$'s were conjectured in \cite{GHM1}.
Very recently, these conjectural values were confirmed in \cite{OZ} by solving TBA-like equations.
Using these values, we can easily translate them into the exact values of the partition function:
\be
\ba
Z_{b^2=3}(1)&=\frac{1}{9}, \qquad\qquad Z_{b^2=3}(2)=\frac{1}{12 \sqrt{3} \pi }-\frac{1}{81}, \\
Z_{b^2=3}(3)&=\frac{5}{2187}-\frac{1}{72 \pi ^2}-\frac{1}{216 \sqrt{3} \pi }, \\
Z_{b^2=3}(4)&=\frac{17}{19683}-\frac{5}{1296 \pi ^2}-\frac{5}{1944 \sqrt{3} \pi }.
\ea
\label{eq:Z3-exact}
\ee
The results for $N=1,2$ are perfectly consistent with the results in the previous section.
See \cite{OZ} for more on the exact values of $\Tr \rho^\ell$ up to $\ell=10$.

One important consequence in \cite{GHM1} is that the large $\mu$ expansion of the
grand partition function in this Fermi-gas system is completely determined by the topological string on local $\mathbb{P}^2$.
In particular, in the current case ($\hbar=2\pi$), the grand partition function can be written
in closed form. Here we show only the result in \cite{GHM1}.
The grand partition function is exactly given by
\be
\Xi_{b^2=3}(\mu)=\re^{J(\mu)}\vartheta_3\( \xi-\frac{3}{8}, \frac{9\tau}{4} \),
\label{eq:Xi-b3}
\ee
where $\vartheta_3(z,\tau)$ is Jacobi's theta function defined by
\be
\vartheta_3(z,\tau):=\sum_{n \in \mathbb{Z}} \re^{\pi \ri n^2 \tau+2\pi \ri n z}.
\ee
All the quantities $J(\mu)$, $\xi$ and $\tau$ in \eqref{eq:Xi-b3} are expressed in terms of the topological
string free energy.
The function $J(\mu)$ is called the modified grand potential,%
\footnote{Obviously, the modified grand potential here is different from the standard grand potential (the logarithm
of the grand partition function), due to the factor of the theta function. As was first observed in \cite{HMO2}, it is more convenient
to consider the modified grand potential rather than the standard grand potential.
This difference is important in identifying an integration contour $\cC$ in \eqref{eq:Z-trans}.
In the standard grand potential, the contour $\cC$ is the finite interval $[-\pi \ri, \pi \ri]$.}
which is written as
\be
\ba
J(\mu)=\frac{1}{4\pi^2} \( F_0(t)-t \pd_t F_0(t)+\frac{t^2}{2} \pd_t^2 F_0(t) \)
+F_1(t)+F_1^\text{NS}(t)-\frac{\zeta(3)}{3\pi^2}+\frac{\log 3}{6}.
\ea
\label{eq:J-top}
\ee
where $F_0(t)$ and $F_1(t)$ are the standard free energies at genus zero and genus one, respectively.
The function $F_1^\text{NS}(t)$ is the first correction to the refined topological string free energy 
in the Nekrasov-Shatashvili limit.
The explicit forms of these functions are shown in Appendix~\ref{sec:localP2}.
The functions $\xi$ and $\tau$ are also related to the genus zero free energy
\be
\xi=\frac{3}{4\pi^2} ( t \pd_t^2 F_0(t)-\pd_t F_0(t)), \qquad
\tau=\frac{2\ri}{\pi} \pd_t^2 F_0(t).
\ee
The K\"ahler modulus $t$ is relate to the chemical potential $\mu$ by the mirror map:%
\footnote{Strictly speaking, the relationship \eqref{eq:mirror-map} is the \textit{quantum} mirror map \cite{ACDKV} for the special case $\hbar=2\pi$.
In this case, the quantum mirror map is essentially same as the standard (classical) mirror map for $\hbar=0$.}
\be
t=-\log z -6z \, {}_4F_3\( 1,1,\frac{4}{3},\frac{5}{3};2,2,2;27z \),\qquad
z=\re^{-3\mu}.
\label{eq:mirror-map}
\ee
Note that the parameter $t$ is related to an ``effective'' chemical potential introduced in \cite{HMO3}.%
\footnote{A notational remark: In \cite{HMO2, HMMO}, the chemical potential was identified as the K\"ahler
parameter. As a result, we had to introduce a new ``effective'' K\"ahler parameter, whose interpretation is unclear, 
as a counterpart of the effective chemical potential $\mu_\text{eff}$.
A more sophisticated identification is to relate $\mu$ to the complex modulus $z$, as in \eqref{eq:mirror-map}.
In this identification, $\mu_\text{eff}$ is naturally related to the K\"ahler modulus,
and we no longer have to introduce the effective modulus. The relation between $\mu_\text{eff}$ and $\mu$ in \cite{HMO3}
is nothing but the quantum mirror map in \cite{ACDKV}. See \cite{CGM1} for example.} 
Plugging the mirror map into the modified grand potential, one finds the following large $\mu$
expansion
\be
\ba
J(\mu)&= \frac{C_3}{3}\mu^3+B_3 \mu+ A_3+  
\left(-\frac{45 \mu ^2}{8 \pi ^2}-\frac{9 \mu }{4 \pi ^2}-\frac{3}{4 \pi ^2}+\frac{3}{8}\right) \re^{-3 \mu} \\
& \quad + 
 \left(-\frac{999 \mu ^2}{16 \pi ^2}-\frac{63 \mu }{16 \pi ^2}+\frac{9}{32} \left(34-\frac{5}{\pi ^2}\right) \right)\re^{-6 \mu} 
 +\cO(\re^{-9\mu}).
\ea
\label{eq:J-large-mu}
\ee
where 
\be
C_3=\frac{9}{8\pi^2},\qquad B_3=\frac{1}{8},\qquad
A_3=- {\zeta(3) \over 3 \pi^2}+ {\log 3 \over 6}.
\label{eq:ABC-3}
\ee
Once the large $\mu$ expansion of $J(\mu)$ is understood, we can know the large $N$ expansion of
the partition function.
The partition function is recovered by the inverse Laplace transform:
\be
Z_{b^2}(N)=\int_{\mathcal{C}} \rd \mu\, \re^{J(\mu)-N\mu},
\label{eq:Z-trans}
\ee
where $\mathcal{C}$ should be chosen as the same contour in the integral representation of the Airy function.
For the expansion \eqref{eq:J-large-mu}, one can write the exponential of $J(\mu)$ as
\be
\re^{J(\mu)}=\re^{J^\text{(p)}(\mu)}\sum_{\ell=0}^\infty \re^{-3\ell \mu} \sum_{n=0}^{2\ell} f_{\ell,n} \mu^n,
\ee
where $J^\text{(p)}(\mu)$ is the cubic polynomial in \eqref{eq:J-large-mu}.
We stress that all the coefficients $f_{\ell, n}$ can be computed by the topological string results \eqref{eq:J-top} and \eqref{eq:mirror-map}.
Thus the large $N$ expansion of the partition function
is generically written as the sum of (the derivatives of) the Airy function:
\be
\ba
Z_{b^2=3}(N)=C_3^{-1/3}\re^{A_3} \sum_{\ell=0}^\infty \sum_{n=0}^{2\ell}
f_{\ell,n} \( -\frac{\pd}{\pd N}\)^n \Ai[ C_3^{-1/3}(N+3\ell-B_3) ],
\ea
\label{eq:Z3-largeN}
\ee
In particular, the leading contribution ($\ell=0$) is
\be
Z_{b^2=3}(N)=C_3^{-1/3}\re^{A_3} \Ai[ C_3^{-1/3}(N-B_3) ] +\cdots.
\label{eq:Z3-pert}
\ee
This Airy functional behavior was first found in \cite{FHM}.
Using the asymptotic expansion of the Airy function at infinity, we finally obtain
\be
-\log Z_{b^2=3}(N)=\frac{2}{3\sqrt{C_3}}N^{3/2}-\frac{B_3}{\sqrt{C_3}}\sqrt{N}+\frac{1}{4}\log N+\cO(1),\qquad
N \to \infty.
\label{eq:Z3-largeN-2}
\ee
The logarithmic term does not depend on the values of $C_3$, $B_3$ and $A_3$, and thus is universal
for 3d Chern-Simons-matter theories with gravity duals.
Such a universal behavior was indeed reproduced by the one-loop supergravity calculation \cite{BGMS}.
The leading $N^{3/2}$ term is
\be
\frac{2}{3\sqrt{C_3}}N^{3/2}=\frac{4 \sqrt{2} \pi }{9}N^{3/2}.
\label{eq:N32-b3}
\ee
Now, we compare this result with the general result in \cite{IY, MPS}.
According to these papers, the $N^{3/2}$ term in ABJM theory with general $k$ and $b$ is
\be
\frac{Q^2}{4}\frac{\pi\sqrt{2k}}{3}N^{3/2}.
\label{eq:N32-general-b}
\ee
For $k=1$ and $b^2=3$, one finds
\be
\frac{Q^2}{4}\frac{\pi\sqrt{2k}}{3}N^{3/2}=\frac{4 \sqrt{2} \pi }{9}N^{3/2}.
\ee
Both results are in perfect agreement.
In \cite{IY, MPS}, the similar formula to \eqref{eq:N32-general-b} was derived for a wide class of Chern-Simons-matter theories on squashed spheres.
Also, it was shown in \cite{MP} that the Airy functional behavior is universal in Chern-Simons-matter theories on $S^3$.
Therefore, it is quite natural to assume that the modified grand potential for general $b$ takes the form
\be
J_{b^2}(\mu)=\frac{C_{b^2}}{3}\mu^3+B_{b^2} \mu+A_{b^2}+\cdots,\qquad \mu \to \infty.
\label{eq:Jb2-large}
\ee
Then, to reproduce \eqref{eq:N32-general-b}, the coefficient $C_{b^2}$ must be
\be
C_{b^2}=\frac{32}{\pi^2 k Q^4}=\( \frac{2}{Q} \)^4 C_1.
\label{eq:Cb2}
\ee
If the next-to-leading term (the $N^{1/2}$-term) is known, one can fix the constant $B_{b^2}$.

The sub-leading contributions ($\ell \geq 1$) in \eqref{eq:Z3-largeN} are non-perturbative
corrections in the $1/N$ expansion, of the form $\re^{-2\pi \ell \sqrt{2N}}$.
The convergence of the sum \eqref{eq:Z3-largeN} is quite rapid.
Let us denote a truncated sum in \eqref{eq:Z3-largeN} at $\ell=\ell_{\text{max}}$ by $Z_{b^2=3}^{\ell_\text{max}}(N)$.
The all-order perturbative $1/N$ resummation \eqref{eq:Z3-pert} gives the value for $N=1$
\be
Z_{b^2=3}^{\ell_\text{max}=0}(1)=0.1111134\dots,
\ee
which is already close to the exact value $Z_{b^2=3}(1)=1/9$.
If taking $\ell_\text{max}=2$, one obtains
\be
Z_{b^2=3}^{\ell_\text{max}=2}(1)=0.1111111111111111111111129\dots.
\ee
For $\ell_\text{max}=10$, the truncated sum for $N=1$ shows an agreement with the exact value with $160$ decimal precision!
Of course, the agreement gets better as $N$ grows.
We conclude that the large $N$ expansion \eqref{eq:Z3-largeN} reproduces the finite $N$ result correctly.

\subsection{A generalization: adding fundamental matters}
Here we give a simple generalization on the mirror side: adding more fundamental hypermultiplets.
Let us consider the $\mathcal{N}=4$
$U(N)$ SYM with one adjoint hypermultiplet and $N_f$ fundamental hypermultiplets.
This theory on $S^3$ was analyzed in detail in \cite{MePu, GM, HO1} (see also for more general setups 
\cite{MN1, MN2, MN3, HHO}) .
The fundamental matter contribution in \eqref{eq:Z-mirror-k=1} is simply replaced by
\be
\widetilde{Z}^{\text{$N_f$-fund}}=\prod_i \cD_b(\lambda_i)^{N_f}.
\ee
It is obvious to see that this replacement changes the density matrix for $b^2=3$ as follows
\be
\rho(x_1,x_2)=\frac{1}{2\pi} \( \frac{\sinh(\frac{x_1}{2})}{\sinh(\frac{3x_1}{2})} \)^{N_f/2}
 \frac{1}{2\cosh (\frac{x_1-x_2}{2}+\frac{\pi \ri}{6})} \( \frac{\sinh(\frac{x_2}{2})}{\sinh(\frac{3x_2}{2})} \)^{N_f/2}.
 \label{eq:rho-Nf}
\ee
This operator can be written as
\be
\ba
\rho(x_1,x_2)&=\frac{1}{\hbar} \bra{x_1} \hat{\rho} \ket{x_2},\qquad \hbar=2\pi, \\
\hat{\rho}&=V(\mx)^{1/2} \frac{\re^{-\mm/6}}{2\cosh \frac{\mm}{2}} V(\mx)^{1/2},
\ea
\ee
where the canonical variables $\mx$ and $\mm$ satisfy the commutation relation
\be
[\mx,\mm]=\ri \hbar.
\ee
Note that the Planck constant $\hbar$ here is different from the one in \eqref{eq:Z-P2}.
In particular, $\hbar$ here is independent of $b$, which we have already set to be $\sqrt{3}$.
In our case, the potential $V$ is
\be
V(x)=\( \frac{\sinh(\frac{x}{2})}{\sinh(\frac{3x}{2})} \)^{N_f}=\frac{1}{(2\cosh x+1)^{N_f}}.
\label{eq:V}
\ee
In the analysis below, it is useful to regard $\hbar$ as a parameter,
and we set $\hbar=2\pi$ at the end.
We want to know the large $N$ behavior for the generalized density matrix \eqref{eq:rho-Nf}.
Unlike the case of $N_f=1$, there seems to be no nice connection to the topological strings,
and thus the analysis is much more difficult.
To explore the large $N$ behavior, we introduce a spectral zeta function by
\be
\zeta_S(s):=\Tr \hat{\rho}^s=\sum_{n=0}^\infty \re^{-s E_n}.
\ee
This function is well-defined for $\real s>0$. For $\real s \leq 0$, it is analytically continued.
As shown in \cite{Hatsuda1}, the grand potential is wirtten as the following Barnes-type integral:
\be
\cJ(\mu):=\log \Xi (\mu)=-\int_{c-\ri \infty}^{c+\ri \infty} \frac{\rd s}{2\pi \ri} \Gamma(s) \Gamma(-s) \zeta_S(s) \re^{s\mu}.
\label{eq:J-Barnes}
\ee
Note that $\cJ(\mu)$ is the standard grand potential, not the modified grand potential.
A constant $c$ must be taken in the range $0<c<1$.

\begin{figure}[tb]
\begin{center}
\resizebox{60mm}{!}{\includegraphics{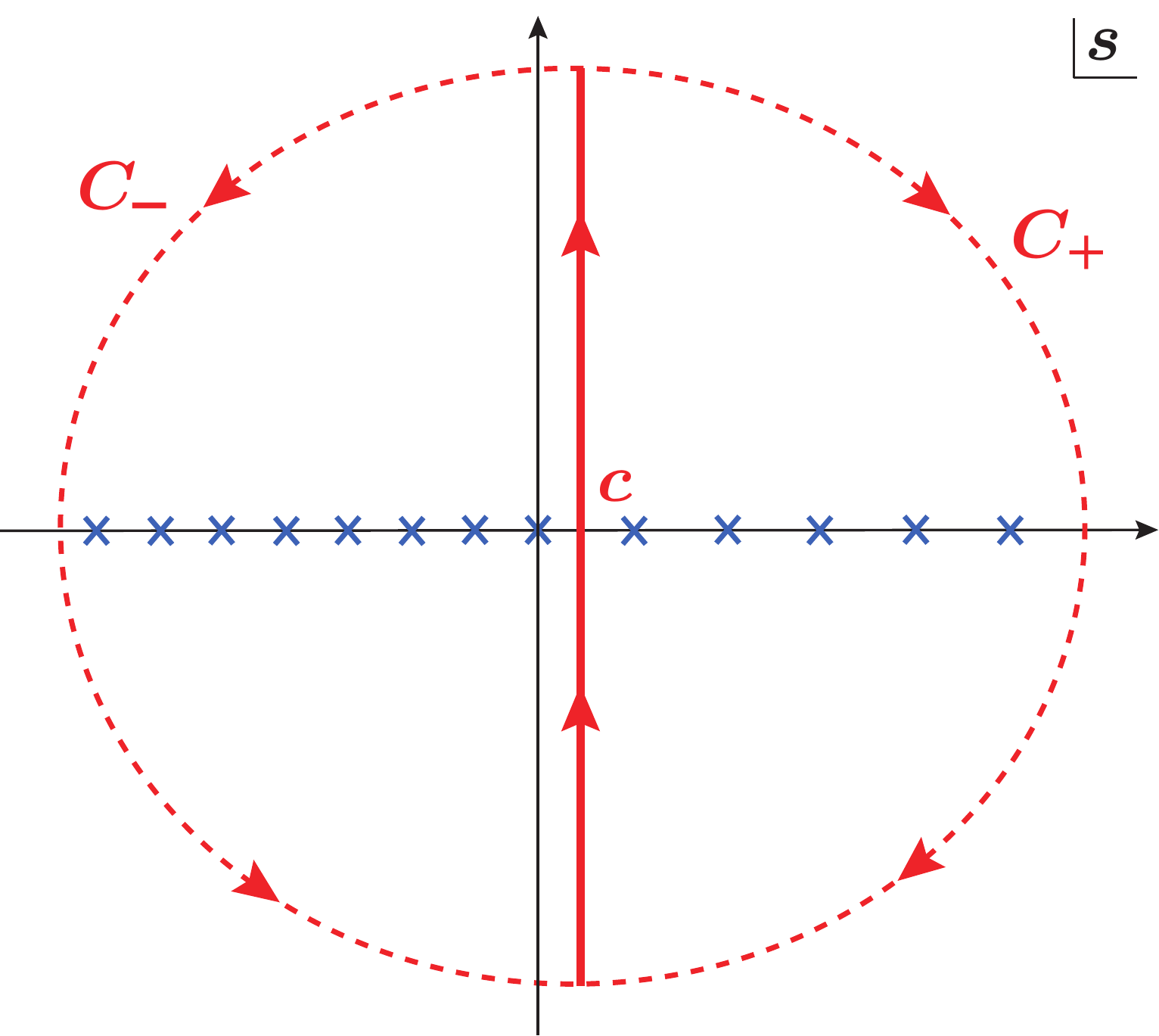}}
\end{center}
\vspace{-0.5cm}
  \caption{One can deform the integration contour in \eqref{eq:J-Barnes} into a closed path by adding an infinite semi-circle
 $C_+$ or $C_-$ for $\mu<0$ or $\mu>0$, respectively. Then, the integral is evaluated by the sum of the residues.
We schematically show the poles of the integrand.}
  \label{fig:contour}
\end{figure}

Let us consider the semi-classical expansion around $\hbar=0$.
In this limit, the grand potential and the spectral zeta admit the WKB expansions:
\be
\cJ^\text{WKB}(\hbar, \mu)=\frac{1}{\hbar} \sum_{n=0}^\infty \hbar^{2n} \cJ^{(n)}(\mu), \qquad
\zeta_S(s)=\frac{1}{\hbar} \sum_{n=0}^\infty \hbar^{2n} \zeta_S^{(n)}(s).
\ee
In the leading approximation, one can treat the canonical variables as classical ones.
Therefore at the leading order, the density operator reduces to
\be
\hat{\rho} \to \rho_\text{cl}(x,p)=V(x)\frac{\re^{-p/6}}{2\cosh \frac{p}{2}}.
\ee
Then, the spectral zeta is computed by a phase space integral:
\be
\zeta_S^{(0)}(s)=\int_{-\infty}^\infty \frac{\rd x \rd p}{2\pi} \rho_\text{cl}(x,p)^s.
\label{eq:zeta0}
\ee
For the potential \eqref{eq:V}, this integral can be exactly performed, and one gets
\be
\zeta_S^{(0)}(s)=\frac{1}{2\pi} \B \( \frac{2s}{3},\frac{s}{3} \) \cI( N_f s),
\ee
where $\B(x,y)=\Gamma(x)\Gamma(y)/\Gamma(x+y)$ is Euler's beta function, and $\cI(\beta)$ is
given by
\be
\cI(\beta):=\int_{-\infty}^\infty \frac{\rd x}{(2\cosh x+1)^\beta}
=\frac{1}{3^{\beta-1/2}} \B\( \beta , \frac{1}{2} \) \, {}_2F_1 \( \frac{1}{2},\frac{1}{2}; \beta+\frac{1}{2};\frac{1}{4} \). 
\ee
A useful property of $\cI(\beta)$ is that it satisfies a recurrence relation
\be
\cI(\beta+2)=\frac{\beta}{3(\beta+1)} \cI(\beta)-\frac{2\beta+1}{3(\beta+1)} \cI(\beta+1).
\ee
This relation is derived by using recurrence relations for the hypergeometric function.
Once we know the analytic form of the spectral zeta function, we can compute the large $\mu$
expansion of the grand potential by the Barnes-type integral \eqref{eq:J-Barnes}.
In the large $\mu$ limit, one can add an infinite semi-circle $C_-$ to the integration contour 
so that the closed path encircles all the poles in the region $\real s <c$ in \eqref{eq:J-Barnes}, 
as shown in Fig.~\ref{fig:contour}.
Since the integral is evaluated by the sum of the residues, it is important to understand the pole structure in the integrand.
For the function \eqref{eq:zeta0}, we observe that the integrand has two kinds of poles in $\real s <c$:
\be
s= -\frac{m}{N_f}, -\frac{3m}{2}, \qquad m=0,1,2,\dots.
\ee
The closest pole from the line $\real s=c$ is $s=0$, and it gives the leading contribution in the large $\mu$ limit.
Computing the residue at $s=0$, we obtain
\be
-\Res_{s=0} \Gamma(s)\Gamma(-s)\zeta_S^{(0)}(s)\re^{s\mu}
=\frac{3}{4\pi N_f}\mu^3+\( \frac{7\pi}{12N_f}-\frac{\pi N_f}{2} \) \mu+A^{(0)}(N_f),
\ee
where the constant part is
\be
A^{(0)}(N_f)=\frac{\zeta(3)}{\pi N_f}+N_f^2 \left[ \frac{\zeta(3)}{2\pi}+\frac{\psi^{(1)}(\frac{1}{3})-\psi^{(1)}(\frac{2}{3})}{4\sqrt{3}} \right].
\ee
Here $\psi^{(m)}(z)=\pd_z^{m+1} \log \Gamma(z)$ are the polygamma functions.
The other poles give exponentially suppressed corrections.
For example, at the pole $s=-1/N_f$, we find the correction of the order $\re^{-\mu/N_f}$,
\be
-\Res_{s=-1/N_f} \Gamma(s)\Gamma(-s)\zeta_S^{(0)}(s)\re^{s\mu}=-\frac{1}{\pi N_f}\Gamma\( \frac{1}{N_f} \) \Gamma\(-\frac{1}{3N_f} \)
\Gamma\(-\frac{2}{3N_f} \) \re^{-\mu/N_f}.
\ee
We do not care about these corrections here, but we stress that all of these
are completely computed by the Branes-type integral.
In summary, the leading semi-classical contribution in the grand potential shows the following large $\mu$ behavior:
\be
\cJ^{(0)}(\mu)=\frac{3}{4\pi N_f}\mu^3+\( \frac{7\pi}{12N_f}-\frac{\pi N_f}{2} \) \mu+A^{(0)}(N_f)+\cO(\re^{-\mu/N_f}, \re^{-3\mu/2}).
\ee

As explained in \cite{MP}, the quantum corrections $\cJ^{(n)}(\mu)$ can be systematically computed by Wigner's method
in phase space. We do not show explicit computations in detail here, since it is straightforward to apply it to our case. 
After a lengthy computation, we obtain
the first correction
\be
\zeta_S^{(1)}(s)=\frac{N_f^2 s^2(s-1)}{216\pi(N_f s+1)} \B\( \frac{2s}{3}, \frac{s}{3} \) \( \cI(N_f s+1) -\cI(N_f s) \).
\ee
From this result, one finds
\be
\cJ^{(1)}(\mu)=\frac{N_f}{24\pi}\mu+A^{(1)}(N_f)+\cO(\re^{-\mu/N_f}, \re^{-3\mu/2}),
\ee
where
\be
A^{(1)}(N_f)=-\frac{N_f}{24\pi}-N_f^2 \( \frac{1}{24\pi}+ \frac{1}{72\sqrt{3}}\).
\ee
The computation of the higher corrections is much more complicated.
In the case of $b=1$, the higher correction starts from the constant term in $\mu \to \infty$.
It is natural to assume this for $b^2=3$:
\be
\cJ^{(n)}(\mu)=A^{(n)}(N_f)+\cO(\re^{-\mu/N_f}, \re^{-3\mu/2}), \qquad n \geq 2.
\ee
Assuming this ansatz, we conclude that the large $\mu$ expansion of the grand potential is generically given by
\be
\cJ_{b^2=3}(\hbar, N_f , \mu)=\frac{3}{4\pi N_f \hbar}\mu^3+\left[ \frac{1}{\hbar}\(\frac{7\pi}{12N_f}-\frac{\pi N_f}{2} \)+\frac{\hbar N_f}{24\pi} \right]\mu
+A_3(\hbar, N_f)+\cdots,
\ee
where
\be
A_3(\hbar, N_f) = \frac{1}{\hbar} \sum_{n=0}^\infty \hbar^{2n} A^{(n)}(N_f).
\label{eq:A3-full}
\ee
A nice property of this result is that the coefficients of $\mu^3$ and $\mu$ do not receive the higher order quantum corrections,
thus one can extrapolate this result to finite $\hbar$.
We are interested in $\hbar=2\pi$, and in this case we finally obtain
\be
\cJ_{b^2=3}(N_f , \mu)=\frac{3}{8\pi^2 N_f}\mu^3+\( \frac{7}{24N_f}-\frac{N_f}{6} \)\mu+A_3(N_f)+\cdots, \qquad
\mu \to \infty.
\label{eq:J3-large-mu}
\ee
For $N_f=1$, it reproduces $C_3$ and $B_3$ in \eqref{eq:ABC-3} correctly.
Since the constant part \eqref{eq:A3-full} receives an infinite number of quantum corrections in the WKB expansion,
we have to resum it for the extrapolation to $\hbar=2\pi$.
In the case of $b=1$, this can be done \cite{HO1}, but in the current case, we do not know its exact form so far.
It is interesting to perform the resummation, and compare the result for $\hbar=2\pi$ and $N_f=1$ with $A_3$
in \eqref{eq:ABC-3}.
Note that the full grand potential receives non-perturbative correction of the form $\re^{-\alpha \mu/\hbar}$, in general. 
To explore these corrections is beyond the scope of this note.
Using the large $\mu$ expansion \eqref{eq:J3-large-mu}, one can easily compute the large $N$ expansion of the partition function.
As in \eqref{eq:Z3-largeN-2}, one finds
\be
-\log Z_{b^2=3}(N_f, N)=\frac{4\pi \sqrt{2N_f}}{9}N^{3/2}-\frac{2\pi \sqrt{2N_f}}{3} \( \frac{7}{24 N_f}-\frac{N_f}{6} \) \sqrt{N}+\frac{1}{4} \log N +\cO(1).
\ee
Let us recall the result for $b=1$ in \cite{MePu}:
\be
-\log Z_{b^2=1}(N_f, N)=\frac{\pi \sqrt{2N_f}}{3}N^{3/2}-\frac{\pi \sqrt{2N_f}}{2} \( \frac{1}{2 N_f}-\frac{N_f}{8} \) \sqrt{N}+\frac{1}{4} \log N +\cO(1).
\ee
These results suggest that the leading $N^{3/2}$-term for general $b$ is given by
\be
\frac{Q^2}{4} \frac{\pi \sqrt{2N_f}}{3}N^{3/2}.
\ee
It is interesting to confirm this directly from the saddle-point analysis in the matrix model.

\section{Supersymmetric R\'enyi entropy}
One interesting application of the ellipsoid partition function is to compute a supersymmetric version of the R\'enyi entropy.
The supersymmetric R\'enyi entropy was proposed in \cite{NY} as a generalization of the R\'enyi entropy.
They share many common properties.
In particular, they reduce to the entanglement entropy in the special limit.
Therefore the supersymmetric R\'enyi entropy plays the role of an order parameter in quantum system.
Originally, the supersymmetric R\'enyi entropy is defined by the partition function on a branched three-sphere with compensating R-symmetry,
called a singular surface in \cite{NY}.
Quite interestingly, in $\cN=2$ supersymmetric gauge theories, the partition function on the singular surface
is exactly equivalent to the one on the ellipsoid \cite{NY}.
As a result, the supersymmetric R\'enyi entropy is finally given by
\be
S_{q}=\frac{1}{1-q} \log \left| \frac{Z_{b^2=q}}{(Z_{b^2=1})^q} \right|.
\label{eq:Sq}
\ee
The entanglement entropy is recovered in the limit $q \to 1$, and it results in $S_1=\log |Z_{b^2=1}|$.
The super R\'enyi entropy was generalized to other dimensions \cite{CDS, ARS, HNU}.
In our case, the ellipsoid partition function is always real-valued, and defining the free energy by $F_{q}=-\log Z_{b^2=q}$,
one can rewrite it as
\be
S_{q}=\frac{1}{1-q}(qF_1-F_q).
\ee
Note that the form \eqref{eq:Sq} manifestly breaks the symmetry in $q \leftrightarrow q^{-1}$,
but since we know $Z_{b^2}=Z_{b^{-2}}$, $S_q$ for $q <1$ is easily obtained from the one for $q >1$.
More explicitly, we have a reflection formula
\be
S_{q^{-1}}=\frac{1}{1-q}(qF_q-F_1)=-q S_q+(1+q)S_1,
\ee
where we used $S_1=-F_1$.
This is useful to analyze the limits $q \to 0$ and $q \to \infty$.
In the limit $q \to \infty$, $S_q$ in general converges to a constant, and thus
$S_q$ in $q \to 0$ is divergent:
\be
S_q=\frac{S_1-S_{\infty}}{q}+\cO(1), \qquad q \to 0,
\ee  
where $S_{\infty}$ is a convergent value of $S_q$ at infinity.
In \cite{NY}, $S_{q}$ in ABJM theory for $N=1$ was analyzed. 
Here we present more results on $S_{q}$ using the results in the previous sections.

As a warm up, let us start with the simplest case $N=1$.
As mentioned above, this case has been analyzed in \cite{NY}, but we give a little bit more precise results.
The exact ellipsoid partition function for $N=1$ is given by \eqref{eq:Z-ABJM-N=1}.
Therefore the exact super R\'enyi entropy is
\be
\ba
S_q^\text{ABJM}(k, N=1)
&=-\log k +\frac{2q \log 2}{1-q}+\frac{2}{1-q} \log \cD_b (0) \\
&=-\log k +\frac{2q \log 2}{1-q}+\frac{4}{1-q} \log s_b \( \frac{\ri Q}{4} \).
\ea
\ee 
For some $q$'s, we find the exact values
\be
\ba
S_1^\text{ABJM}(k, N=1)&=-\log(4k), \\
S_2^\text{ABJM}(k, N=1)&=-\log(4k)+\frac{\mathrm{K}}{\pi}-\frac{7\log 2}{4}+\log(2+\sqrt{2}), \\
S_3^\text{ABJM}(k, N=1)&=-\log(4k)+\log \( \frac{3}{2} \), \\
S_5^\text{ABJM}(k, N=1)&=-\log(4k)+\frac{1}{2} \log \( \frac{3+\sqrt{5}}{2} \),
\ea
\ee
where $\mathrm{K}=0.915965594177\dots$ is Catalan's constant.
Using the result \eqref{eq:Db-b=1-expand}, one finds the expansion around $q=1$,
\be
S_q^\text{ABJM}(k, N=1)=-\log(4k)+\frac{\pi^2}{16}(q-1)-\frac{\pi^2}{16}(q-1)^2+\cO((q-1)^3).
\ee
Also using \eqref{eq:sb-WKB}, one obtains the expansion of the double sine function at $z=\ri Q/4$ around $b=0$,
\be
\log s_b \( \frac{\ri Q}{4} \) = -\frac{\mathrm{K}}{2\pi b^2}-\frac{\log 2}{8}-\frac{\pi b^2}{96}-\frac{\pi^2 b^4}{64}+\cO(b^6),
\ee
Thus the expansion around $q=0$ is
\be
S_q^\text{ABJM}(k, N=1)=
-\frac{2 \mathrm{K}}{\pi  q}-\log k-\frac{2 \mathrm{K}}{\pi }-\frac{\log (2)}{2}+\cO(q), \qquad q \to 0.
\ee
Similarly, the expansion around $q=\infty$ is
\be
S_q^\text{ABJM}(k, N=1)=-\log (4k) +\frac{2 \mathrm{K}}{\pi }+\( -\frac{2 \mathrm{K}}{\pi }
-\frac{3\log 2}{2} \) q^{-1}+ \cO(q^{-2}),\quad q \to \infty
\ee
All of these results are in perfect agreement with the ones in \cite{NY}.%

\begin{figure}[tb]
\begin{center}
\resizebox{80mm}{!}{\includegraphics{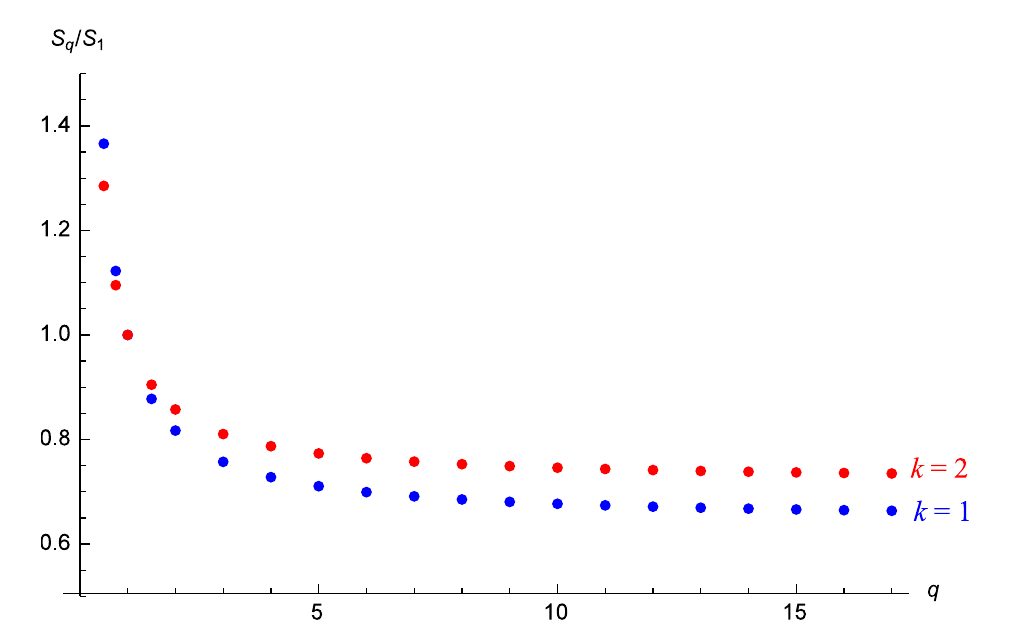}}
\end{center}
\vspace{-0.5cm}
  \caption{The $q$-dependence of the normalized super R\'enyi entropy $\cS_q=S_q/S_1$ 
  for $N=2$ ABJM at $k=1,2$ is shown. 
The blue is the case of $k=1$, and the red is $k=2$.}
  \label{fig:Sq-N2}
\end{figure}

The computation for $N \geq 2$ is much more involved.
In these cases, only the numerical evaluation is available so far.
For $N=2$, we use the integral representation \eqref{eq:ZABJM-2}.
In the round sphere case, the exact partition function was computed in \cite{Okuyama1}:
\be
Z_{b^2=1}^\text{ABJM}(k,N=2)=\frac{1}{16}\int_{-\infty}^\infty \rd \lambda \,
\frac{\lambda \tanh^2 \pi \lambda}{\sinh \pi k \lambda}.
\ee
For integral $k$, this integral can be performed \cite{Okuyama1}.
For $k=1,2$, we have
\be
Z_{b^2=1}^\text{ABJM}(k=1,N=2)=\frac{1}{16\pi}, \qquad
Z_{b^2=1}^\text{ABJM}(k=2,N=2)=\frac{1}{32\pi^2}.
\ee 
Using these, we can evaluate $S_q^\text{ABJM}(k, N=2)$ numerically.
In Fig.~\ref{fig:Sq-N2}, we show the normalized entropy $\cS_q :=S_q/S_1$
for $k=1,2$ as a function of $q$.
As expected in \cite{NY}, these monotonically decrease. 
Using the numerical data, one can estimate the convergent value at $q=\infty$.  
From the numerical fitting by an ansatz $\cS_q=\cS_\infty+\cS_\infty^{(1)}/q+\cS_\infty^{(2)}/q^2+\cdots$, we find
\be
\ba
\cS_\infty(k=1, N=2)  \approx 0.6450, \qquad
\cS_\infty(k=2, N=2)  \approx 0.7193,
\ea
\ee
and the convergent values of the unnormalized entropy:
\be
\ba
S_\infty(k=1, N=2)  \approx -2.527, \qquad
S_\infty(k=2, N=2)  \approx -4.140.
\ea
\ee
It would be interesting to derive these values analytically.

Let us proceed to the large $N$ expansion of $S_q$.
If we assume that the large $\mu$ behavior of the grand potential is given by the form \eqref{eq:Jb2-large},
the free energy is written in terms of the Airy function.
The large $N$ expansion of the free energy is thus
\be
F_q(N)=\frac{2}{3\sqrt{C_q}}N^{3/2}-\frac{B_q}{\sqrt{C_q}}\sqrt{N}+\frac{1}{4}\log N+\cO(1),\qquad
N \to \infty.
\label{eq:Fq-largeN}
\ee
Then the large $N$ expansion of $S_q(N)$ is generically written as
\be
S_q(N)=\frac{2}{3(1-q)}\( \frac{q}{\sqrt{C_1}}-\frac{1}{\sqrt{C_q}} \) N^{3/2}
-\frac{1}{1-q} \(\frac{q B_1}{\sqrt{C_1}}-\frac{B_q}{\sqrt{C_q}} \)\sqrt{N}-\frac{1}{4}\log N+\cO(1).
\ee
As in the free energy, the logarithmic term is universal.
Using the results in \cite{IY, MPS}, we have \eqref{eq:Cb2}.
Therefore the $N^{3/2}$-term is
\be
-\frac{2}{3\sqrt{C_1}} \frac{3q+1}{4q} N^{3/2}.
\ee
This is of course consistent with the result in \cite{NY}.
We do not know the explicit form of $B_q$ except for $q=1,3$.
As seen in the previous section, in the generalized mirror theory with $N_f$ fundamental hypers, we have
\begin{alignat}{3}
C_1(N_f)&=\frac{2}{\pi^2 N_f}, &\qquad B_1(N_f)&=\frac{1}{2N_f}-\frac{N_f}{8}, \\
C_3(N_f)&=\frac{9}{8\pi^2 N_f}, &\qquad B_3(N_f)&=\frac{7}{24N_f}-\frac{N_f}{6}.
\end{alignat}
The large $N$ expansion of the third super R\'enyi entropy $S_3$ in this theory thus reads
\be
S_3(N_f,N)=-\frac{5\pi \sqrt{N_f}}{9\sqrt{2}} N^{3/2}
-\frac{\pi(11N_f^2-80)}{144 \sqrt{2N_f}} \sqrt{N}-\frac{1}{4} \log N+\cO(1).
\ee

\begin{figure}[tb]
\begin{center}
\begin{tabular}{cc}
\resizebox{65mm}{!}{\includegraphics{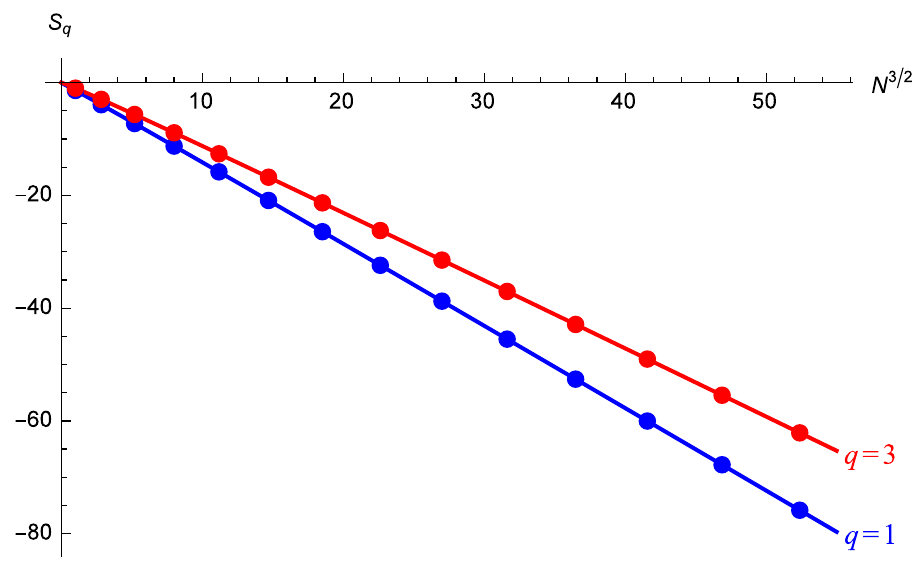}}
\hspace{6mm}
\resizebox{65mm}{!}{\includegraphics{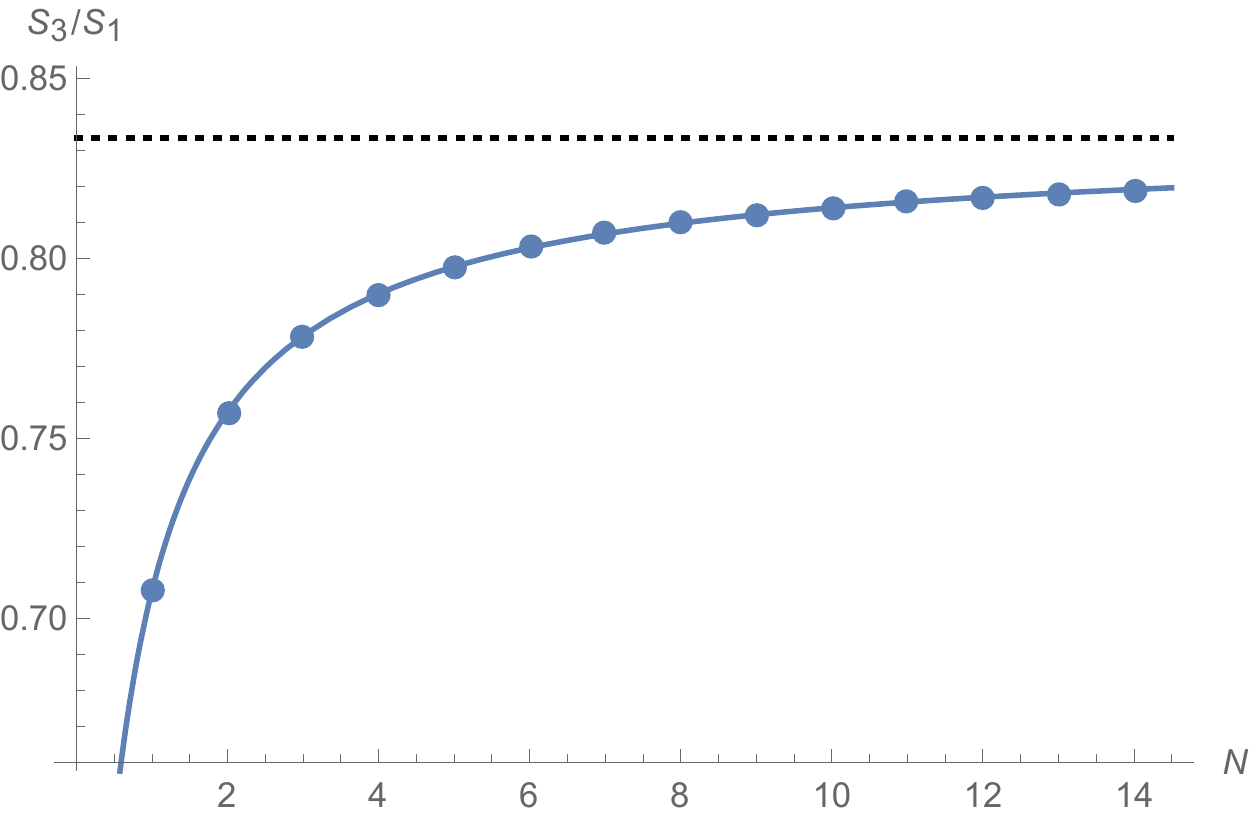}}
\end{tabular}
\end{center}
  \caption{(Left) The first and third super R\'enyi entropies for the mirror ABJM are plotted against $N^{3/2}$. (Right) The normalized
third entropy $\cS_3$ is shown as a function of $N$. In both figures, the dots represent the exact values, and the solid lines are the leading Airy contributions.}
  \label{fig:S3}
\end{figure}

Finally, for $N_f=1$, one can compute the exact values of $Z_{b^2=3}(N_f=1,N)$ for various $N$ (see \eqref{eq:Z3-exact}).%
\footnote{Using the technique in \cite{OZ}, one can compute the exact values for various values of $N_f$.}
The exact values of $Z_{b^2=1}(N_f=1,N)$ also have been computed in \cite{HMO1, PY} (see also \cite{CGM1}).
Therefore we can compute the exact values of $S_3(N_f=1,N)$.
For $b^2=1$, the leading Airy functional form is
\be
Z_{b^2=1}(N_f=1,N)=C_1^{-1/3}\re^{A_1} \Ai[ C_1^{-1/3}(N-B_1) ] +\cdots.
\label{eq:Z1-pert}
\ee
where
\be
C_1=\frac{2}{\pi^2},\qquad B_1=\frac{3}{8},\qquad A_1=\frac{\log 2}{4}-\frac{\zeta(3)}{8\pi^2}.
\ee
We show, in the left of Fig.~\ref{fig:S3}, the behaviors of $S_1(1,N)$ and $S_3(1,N)$ against $N^{3/2}$.
The dots are the exact values computed by the Fermi-gas formalism in the previous section,%
\footnote{We thank Szabolcs Zakany for sharing his result on the exact spectral trace $\Tr \rho^\ell$
up to $\ell=14$ for local $\mathbb{P}^2$ with $\hbar=2\pi$.}
while the solid lines represent the results obtained by using the leading large $N$ contribution \eqref{eq:Z3-pert} and \eqref{eq:Z1-pert}.
We can see that $S_3(1,N)$ indeed scales as $N^{3/2}$.
Also in the right of Fig.~\ref{fig:S3}, we plot the normalized entropy $\cS_3(1,N)$.
It is getting closer to the expected value $(3q+1)/(4q)=5/6$ at large $N$.

\section{Concluding remarks}
In this note, we studied the ellipsoid partition functions in ABJM theory and in its mirror dual.
In the special case of $b^2=3$, the mirror matrix model \eqref{eq:Z-mirror} has an unexpected connection to the topological string on local $\mathbb{P}^2$.
This surprising connection allows us to compute the large $N$ expansion including all the non-perturbative corrections in $1/N$.
This is the first result of the large $N$ solution in ABJM theory for $b \ne 1$.
In a sense, our result provides a ``field theoretic realization'' of the topological string on local $\mathbb{P}^2$ with the special string coupling
$g_s=1/\hbar=(2\pi)^{-1}$.
Using the obtained results, we analyzed the supersymmetric R\'enyi entropy.

There are several points to be clarified in the future.
Here we mainly focused on the special case $b^2=3$. Needless to say, it is important to understand the structure
at large $N$ for general $b$.
The remarkable fact found in \cite{FHM, MP} is that in 3d Chern-Simons-matter theories that have AdS$_4$ gravity duals,
the all-order perturbative $1/N$ expansions are resummed as the universal form \eqref{eq:Z3-pert} in terms of the Airy function.
In other words, all the information on the perturbative $1/N$ corrections is encoded in only the three constants $A$, $B$ and $C$
of the cubic polynomial in the grand potential. 
Since the constant $A$ appears as an overall factor of the partition function, it is not relevant
in the $1/N$ expansion (but important in the comparison with the exact data).
In this note, we fixed the constant $C$ for general $b$, as in \eqref{eq:Cb2}, by comparing the known results in \cite{IY, MPS}.
It is very important to fix $B$ as a function of $b$.
For this goal, it is sufficient to compute the $N^{1/2}$ term in the free energy.
The analysis in \cite{MePu} and its refinement would be useful.

For generic values of $b$, the evaluation of the matrix integrals is very difficult.
If $b^2$ is an odd integer, the matrix model also simplifies, thus these cases are good examples as a next step.
Also, the expansion around $b=1$ is interesting. Near $b=1$, the function $\cD_b(\lambda)$ is
written in terms of the hyperbolic functions, as in \eqref{eq:Db-b=1-expand}.
A more direct approach to evaluate the matrix integrals is to use Monte Carlo methods.
In ABJM theory on the round sphere, this approach is greatly successful \cite{KEK}, and we expect that this also works for
the mirror matrix model \eqref{eq:Z-mirror} for general $b$ \cite{HHY}.%
\footnote{As mentioned in \cite{KEK}, in the Monte Carlo study, the mirror matrix model is much more suitable than the original ABJM
matrix model due to a sign problem. However, it is an interesting problem to put the ABJM matrix model on the Monte Carlo method
directly, because, if this is possible, we can discuss 3d mirror symmetry for relatively larger $N$.}  

It would be interesting to explore a factorization property of the ellipsoid partition functions \cite{Pasquetti}.
Though it seems almost impossible to perform the matrix integral \eqref{eq:Z-ABJM} or \eqref{eq:Z-mirror} directly,
it might be useful to use the Higgs branch localization \cite{FHY, BP}.

Since ABJM theory is a fundamental prototype in theories on multiple M2-branes,
there are a huge number of generalizations.
One simple direction is to analyze the ABJ partition function.
The main difficulty in this case is that the physical interpretation of the mirror theory is unclear.
In the case of $b=1$, one can directly rewrite the ABJ matrix model as the form of the ``mirror'' partition function \cite{AHS}
by using the Cauchy determinant formula \cite{Honda1}. However for $b \ne 1$, we cannot immediately use this formula,
and it is hard to find such a mirror description.
This difficulty is closely related to proving 3d mirror symmetry \eqref{eq:mirror-equal-k=1}.
It is desirable to understand 3d mirror symmetry more deeply.
For ABJ theory with $b=1$, the grand canonical partition function satisfies beautiful functional relations \cite{GHM2} (see also \cite{MaMo, HoOk}), 
similar to the so-called quantum Wronskian. It is natural to ask whether such functional relations exist for $b \ne 1$ or not.
Finally it is very significant to understand instanton effects from the dual gravity perspective.

\acknowledgments{
I would like to thank Santiago Codesido, Sebasti\'an Franco, Alba Grassi, Masazumi Honda, Marcos Mari\~no, 
Sanefumi Moriyama, Tomoki Nosaka, Kazumi Okuyama,  Julian Sonner, J\"org Teschner, Daisuke Yokoyama, 
Shuichi Yokoyama and Szabolcs Zakany for useful discussions. 
I am especially grateful to Marcos Mari\~no for helpful comments on the manuscript.
This work is supported in part by the Fonds National Suisse, subsidies 200021-156995 and by the NCCR 51NF40-141869 
“The Mathematics of Physics” (SwissMAP).
}

\appendix

\section{The double sine function and related functions}\label{sec:double-sine}
The double sine function is defined by
\be
s_b(z):=\prod_{m,n=0}^\infty \frac{mb+nb^{-1}+\frac{Q}{2}-\ri z}{mb+nb^{-1}+\frac{Q}{2}+\ri z},
\qquad Q=b+b^{-1}.
\label{eq:double-sine}
\ee
It turns out that this function is closely related to Faddeev's (non-compact) quantum dilogarithm
\be
s_b(z)=\exp \left[ -\frac{\pi \ri}{2} z^2-\frac{\pi \ri}{24}(b^2+b^{-2}) \right] \Phi_b(z)
=\frac{\Phi_b(z)}{\Phi_b(0)}\re^{-\pi \ri z^2/2},
\ee
Here we define the quantum dilogarithm by the following integral representation
\be
\Phi_b(z):=\exp \left[ \int_{\mathbb{R}+\ri 0} \frac{\rd t}{t} \frac{\re^{-2\ri t z}}{4\sinh(b t) \sinh( b^{-1} t)} \right]. 
\label{eq:q-dilog}
\ee
The double sine function is normalized as $s_b(0)=1$, and satisfies several functional equations
\be
\ba
s_b(z)&=s_{b^{-1}}(z) ,\\
s_b(z) s_b(-z)&=1, \\
\overline{s_b(z)}&=s_b(-\bar{z}) ,\\
\frac{s_b(z+\frac{\ri}{2}b^{\pm 1})}{s_b(z-\frac{\ri}{2}b^{\pm 1})}&=\frac{1}{2\cosh (\pi b^{\pm 1} z)}.
\ea
\ee
For $b=1$, $s_{b}(z)$ is written in terms of the classical polylogarithms
\be
s_{b=1}(z)=\exp \left[-\frac{\pi \ri}{2}z^2-\frac{\pi \ri}{12}+\ri z \log(1-\re^{2\pi z})
+\frac{\ri}{2\pi}\Li_2(\re^{2\pi z}) \right].
\ee
The ``free energy'' of the double sine function is expanded in $z \to + \infty$
\be
\ri \log s_b(z)=-\frac{\pi z^2}{2}-\frac{\pi}{24}(b^2+b^{-2})
+\sum_{\ell=1}^\infty \frac{(-1)^{\ell-1}}{\ell} \left[ \frac{\re^{-2\pi \ell b z}}{2\sin(\pi \ell b^2)}
+\frac{\re^{-2\pi \ell z/b}}{2\sin(\pi \ell/b^2)} \right].
\label{eq:sb-large-z}
\ee
This expansion is obtained by rewriting the integral in \eqref{eq:q-dilog} as the sum of the residues.
In the ``semi-classical'' limit $b \to 0$ (with $b z$ kept fixed), the free energy is also expanded as
\be
\ri \log s_b(z) \sim -\frac{\pi z^2}{2}-\frac{\pi}{24}(b^2+b^{-2})-\sum_{g=0}^\infty \frac{(-1)^g B_{2g}(1/2)}{(2g)!}
\Li_{2-2g}(-\re^{-2\pi b z})(2\pi b^2)^{2g-1},
\label{eq:sb-WKB}
\ee
where $B_{2g}(x)$ are Bernoulli polynomials, and $\sim$ means that both hand sides asymptotically equal.

The residues in \eqref{eq:sb-large-z} come from two kinds of poles $t=-\pi \ri \ell b$ and $t=-\pi \ri \ell/b$.
In the limit $b \to 0$, the latter goes to infinity, and is not visible in the semi-classical expansion.
In fact, the pole $t=-\pi \ri \ell/b$ leads to the term of order $\re^{-2\pi \ell z/b}$, which is regarded as
a non-perturbative correction in the semi-classical limit $b \to 0$.%
\footnote{This does not mean the perturbative expansion \eqref{eq:sb-WKB} around $b=0$ is non-Borel summable.
As observed in \cite{HO2}, the asymptotic expansion \eqref{eq:sb-WKB} is \textit{Borel summable} for $b \in \mathbb{R}$
and $z>0$, and interestingly the Borel resummation of \eqref{eq:sb-WKB} reproduces the exact function $s_b(z)$ for finite $b$.
In other words, the S-dual structure under $b \leftrightarrow b^{-1}$ in \eqref{eq:sb-large-z} is recovered only 
by the Borel resummation of \eqref{eq:sb-WKB}.}.
Physically, the contributions from these two poles are interpreted as vortex and anti-vortex free energies \cite{Pasquetti}.
A very similar phenomenon was found in the ABJM Fermi-gas system \cite{Hatsuda1}.
In this system, perturbative and non-perturbative corrections to the grand potential
in the semi-classical limit are also caused by two kinds of poles of the integrand in the Barnes-type integral \eqref{eq:J-Barnes}, 
and these poles naturally lead to
membrane/worldsheet instanton corrections.

In \cite{BT}, a function $D_\alpha(x)$ related to $s_b(z)$ was introduced by
\be
D_\alpha(x):= \frac{s_b(x-\alpha)}{s_b(x+\alpha)}.
\ee
Obviously, the function $\cD_b(\lambda)$ defined in \eqref{eq:Db-def} is the special case of $D_\alpha(\lambda)$ as
\be
\cD_b(\lambda)=D_{\alpha=-\ri Q/4}(\lambda).
\ee
For basic properties of $D_\alpha(x)$, see Appendix~A.3 in \cite{BT}.
It turns out that $\cD_b(\lambda)$ has the following integral representation
\be
\cD_b(\lambda)=\exp \left[ \int_{\mathbb{R}+\ri 0} \frac{\rd t}{t} \frac{\sinh ( \frac{Qt}{2}) \cos (2\lambda t)}
{2\sinh(b t) \sinh( b^{-1} t)} \right]
\ee
From this representation, it is obvious to see
\be
\ba
\cD_b(-\lambda)=\cD_b(\lambda),\qquad
\cD_{b^{-1}}(\lambda)=\cD_b(\lambda).
\ea
\ee
Moreover, $\cD_b(\lambda)$ is real and positive for $\lambda \in \mathbb{R}$:
\be
\cD_b(\lambda)=s_b\( \frac{\ri Q}{4}+\lambda \) s_b\( \frac{\ri Q}{4}-\lambda\)=\left| s_b\( \frac{\ri Q}{4}+\lambda\) \right|^2 >0.
\ee
In the limit $b \to 1$, it reduces to
\be
\cD_{b=1}(\lambda)=\frac{s_{b=1}(\lambda+\frac{\ri}{2})}{s_{b=1}(\lambda-\frac{\ri}{2})}=\frac{1}{2\cosh(\pi \lambda)}.
\ee
More precisely, $\cD_b(\lambda)$ has the following expansion around $b=1$:
\be
\cD_b(\lambda)=\frac{1}{2\cosh(\pi \lambda)}\left[ 1-(b-1)^2 F(\lambda)+(b-1)^3 F(\lambda)+\cO((b-1)^4) \right], \quad
b \to 1,
\label{eq:Db-b=1-expand}
\ee
where
\be
F(\lambda):=\frac{\pi^2(1+4\lambda^2)}{8\cosh^2(\pi \lambda)}+\frac{\pi \lambda}{2}\tanh (\pi \lambda).
\ee
Interestingly, for $b=\sqrt{3}$, the function drastically simplifies,
\be
\cD_{b=\sqrt{3}}(\lambda)=\frac{\sinh ( \frac{\pi}{\sqrt{3}} \lambda)}{\sinh ( \sqrt{3}\pi \lambda)}.
\label{eq:Db-sqrt3}
\ee
Also, for $b=\sqrt{2}$, we find the following complicated expression:
\be
\ba
&\cD_{b=\sqrt{2}}(\lambda)=\frac{1}{2^{1/4}(2\cosh(2\sqrt{2}\pi \lambda))^{1/8}(\sqrt{2}\cosh(\sqrt{2}\pi \lambda)+1)^{1/2}} \\
&\quad \times \exp \left[ -\sqrt{2} \lambda \arctan (\re^{-2\sqrt{2}\pi \lambda})
+\frac{\ri}{4\pi} \( \Li_2( \ri \re^{-2\sqrt{2}\pi \lambda}) -\Li_2( -\ri \re^{-2\sqrt{2}\pi \lambda}) \) \right].
\label{eq:Db-sqrt2}
\ea
\ee
This expression is obtained by performing the sum for $b=\sqrt{2}$ in \eqref{eq:Db-large} exactly.
In particular, for $\lambda=0$, one obtains the exact value
\be
\cD_{\sqrt{2}}(0)=\frac{1}{(8+6\sqrt{2})^{1/4}} \exp \left( -\frac{\mathrm{K}}{2\pi} \right)=0.4289574975659\dots,
\ee
where $\mathrm{K}$ is Catalan's constant.
As in the double sine function, the ``free energy" of $\cD_b(\lambda)$ has the following expansion
\be
\log \cD_b(\lambda)=-\frac{\pi Q}{2} \lambda+\sum_{\ell=1}^\infty
\left[ \frac{\re^{-2\pi \ell b \lambda}}{2\ell \cos (\frac{\pi \ell b}{2}Q)}
+\frac{\re^{-2\pi \ell \lambda/b}}{2\ell \cos (\frac{\pi \ell b^{-1}}{2}Q)}
\right],\qquad \lambda \to + \infty.
\label{eq:Db-large}
\ee
Interestingly, $\cD_b(\lambda)$ is self-dual in the Fourier transform (see (A.29) in \cite{BT}):
\be
\int_{-\infty}^\infty \rd x\,  \re^{2\pi \ri x y}\, \cD_b(x)=\cD_b(y).
\label{eq:Db-Fourier}
\ee

\section{Explicit computations for $N=1,2$}\label{sec:explicit}
In this appendix, we show explicit computations of the partition functions for $N=1,2$.
For $N=1$, the ABJM partition function reduces to
\be
\ba
Z_{b^2}^\text{ABJM}(k, 1)&=\int_{-\infty}^\infty \rd \sigma \rd \tilde{\sigma}
\, \re^{\pi \ri k(\sigma^2-\tilde{\sigma}^2)} \cD_b(\sigma-\tilde{\sigma})^2 \\
&=\int_{-\infty}^\infty \rd \sigma \rd \tilde{\sigma} \, \re^{\pi \ri k(2\sigma \tilde{\sigma}-\tilde{\sigma}^2)}
\cD_b(\tilde{\sigma})^2.
\ea
\ee
The integral over $\sigma$ is just the Fourier transform of the constant:
\be
\int_{-\infty}^\infty \rd \sigma \, \re^{2\pi \ri k \sigma \tilde{\sigma}}=\delta(k \tilde{\sigma}). 
\ee
Thus we finally obtain
\be
\ba
Z_{b^2}^\text{ABJM}(k, 1)=\int_{-\infty}^\infty \rd \tilde{\sigma} \, \delta( k \tilde{\sigma}) 
\re^{-\pi \ri k \tilde{\sigma}^2} \cD_b(\tilde{\sigma})^2 
=\frac{1}{k}\cD_b(0)^2.
\label{eq:Z-ABJM-N=1}
\ea
\ee
On the other hand, the mirror partition function for $N=1$ is computed as
\be
\ba
Z_{b^2}^\text{Mirror}(1)&=\cD_b(0) \int_{-\infty}^\infty \rd \lambda \, \cD_b(\lambda)=\cD_b(0)^2.
\ea
\ee
where we used \eqref{eq:Db-Fourier}.

Next, let us rewrite the matrix integral of the ABJM partition function for $N=2$.
We first change the integration variables by
\be
x=\sigma_1-\sigma_2, \qquad y=\sigma_1-\tilde{\sigma}_1,\qquad z=\sigma_2-\tilde{\sigma}_2.
\ee
Then the matrix integral becomes
\be
\ba
Z_{b^2}^\text{ABJM}(k,2)&=\frac{1}{4}\int \! \rd x \rd y \rd z \rd \sigma_2 \,
\re^{2\pi \ri k(xy+(y+z)\sigma_2)-\pi \ri k (y^2+z^2)} \\
&\quad \times\cD_b(y)^2 \cD_b(z)^2 \cD_b(x-y)^2 \cD_b(x+z)^2 \\
&\hspace{-1cm}\times 4\sinh(\pi b x)\sinh(\pi b^{-1}x)\cdot 4\sinh(\pi b (x-y+z))\sinh(\pi b^{-1}(x-y+z)).
\ea
\ee
The integral over $\sigma_2$ can be performed, and it leads to the delta function $\delta(k(y+z))$.
After performing the integral over $z$, one obtains
\be
\ba
Z_{b^2}^\text{ABJM}(k,2)&=\frac{1}{4k}\int \! \rd x \rd y\, \re^{2\pi \ri k x y-2\pi \ri k y^2} \cD_b(y)^4 \cD_b(x-y)^4 \\
&\times 4\sinh(\pi b x)\sinh(\pi b^{-1}x)\cdot 4\sinh(\pi b (x-2y))\sinh(\pi b^{-1}(x-2y)).
\ea
\ee
Shifting the varable $x \to x+y$, one finally obtains \eqref{eq:ZABJM-2}.

The mirror partition function for $N=2$ is
\be
\ba
Z_{b^2}^\text{Mirror}(2)&=\frac{\cD_b(0)^2}{2} \int\! \rd \lambda_1 \rd \lambda_2 \, \cD_b(\lambda_1)\cD_b(\lambda_2) \\
&\quad \times 4\sinh(\pi b (\lambda_1-\lambda_2))\sinh(\pi b^{-1} (\lambda_1-\lambda_2))\cD_b(\lambda_1-\lambda_2)^2 \\
&=\frac{\cD_b(0)^2}{2} \int\! \rd \lambda_1 \rd \lambda_2 \, \cD_b(\lambda_1)\cD_b(\lambda_1-\lambda_2) \\
&\quad\times 4\sinh(\pi b \lambda_2)\sinh(\pi b^{-1} \lambda_2)\cD_b(\lambda_2)^2 \\
\ea
\ee
Using the Fourier transform \eqref{eq:Db-Fourier}, the convolution part is computed as
\be
\int\! \rd \lambda_1 \cD_b(\lambda_1) \cD_b(\lambda_1-\lambda_2)=\int \! \rd x\,  \re^{2\pi \ri x \lambda_2} \cD_b(x)^2
\ee
Therefore one finally finds the representation \eqref{eq:Zmirror-2}.

\section{The free energy for local $\mathbb{P}^2$}\label{sec:localP2}
To compute the grand partition function in the case of $b^2=3$, we need the topological string free energy for local $\mathbb{P}^2$.
In general, the free energy in the refined topological string has two couplings $\epsilon_1$ and $\epsilon_2$, 
and it admits the following perturbative expansion (see \cite{HKRS}, for instance)
\be
F^\text{ref}(\bm{t};\epsilon_1,\epsilon_2)=\sum_{n,g=0}^\infty (\epsilon_1+\epsilon_2)^{2n}(\epsilon_1 \epsilon_2)^{g-1} F^{(n,g)}(\bm{t}).
\label{eq:F-ref}
\ee
where $\bm{t}=(t_1,t_2,\dots)$ are K\"ahler moduli of the geometry.
The free energies appearing in \eqref{eq:J-top} are related to the coefficients in \eqref{eq:F-ref} by
\be
F_0(t)=F^{(0,0)}(t),\qquad F_1(t)=F^{(0,1)}(t),\qquad F_1^\text{NS}(t)=F^{(1,0)}(t).
\ee
The genus zero free energy (or the prepotential) $F_0(t)$ is computed by solving the Picard-Fuchs equation.
For local $\mathbb{P}^2$, the two basic periods at large radius point are given by, 
\be
\ba
\widetilde \varpi_1(z)&= \sum_{j=1}^\infty 3 {(3j-1)! \over (j!)^3}  z^j
=6z \, {}_4F_3\( 1,1,\frac{4}{3},\frac{5}{3};2,2,2;27z \) , \\
\widetilde \varpi_2(z)&=\sum_{j=1}^\infty 18 
{ (3j-1)!
\over (j!)^3} \left( \psi(3j) - \psi (j+1)\right)z^{j}, 
\ea
\ee
where $\psi(z)=\psi^{(0)}(z)$ is the digamma function.
Then, $F_0(t)$ is implicitly given by
\be
\ba
t&=-\log z-\widetilde \varpi_1(z), \\
\pd_t F_0(t)&= {1\over 6} \left( \log^2(z) + 2  \widetilde \varpi_1(z) \log(z) +  \widetilde \varpi_2(z) \right).
\ea
\ee
Eliminating $z$, one finds
\be
F_0(t)=\frac{t^3}{18}-3\re^{-t}-\frac{45}{8}\re^{-2t}-\frac{244}{9} \re^{-3t}-\frac{12333}{64}\re^{-4t} +\cO(\re^{-5t}).
\ee
The functions $F_1(t)$ and $F_1^\text{NS}(t)$ are written in closed forms:
\be
\ba
F_1(t)&=-\frac{1}{2} \log \( -\frac{\rd t}{\rd z} \)-\frac{1}{12} \log \( z^7(1-27z) \), \\
F_1^\text{NS}(t)&=-\frac{1}{24}\log \( \frac{1-27z}{z} \).
\ea
\ee
Using these results, one can compute the large $\mu$ expansion of $J(\mu)$ up to any desired order.

\end{document}